\documentclass[aps,prb,twocolumn,keywords,nopacs,amssymb]{revtex4}

\usepackage[dvips]{graphicx}
\usepackage{dcolumn}% Align table columns on decimal point
\usepackage{bm}% bold math
\usepackage{color}
\usepackage{bm}% bold math
\usepackage{color}
\def\be{\begin{equation}}
\def\en{\end{equation}}
\def\bea{\begin{eqnarray}}
\def\ena{\end{eqnarray}}
\def\bec{\begin{equation}\begin{array}{rcl}}

\def\p{\partial}

\def\ls{\lesssim}
\def\gs{> \kern -13pt \lower 5pt \hbox{$\displaystyle{\sim}$}}
\def\ls{< \kern -13pt \lower 5pt \hbox{$\displaystyle{\sim}$}}

\def\ve{\varepsilon}
\newcommand{\av}[1]{\langle{#1}\rangle}
\newcommand{\AV}[1]{\bigg \langle{#1}\bigg \rangle}
\newcommand{\bi}[1]{\mbox{\boldmath$#1$}}

\def\a1{\stackrel{\leftrightarrow}{1}}

\begin{document}
\title{Molecular Dynamics Simulation of 
Water between Metal Walls under Electric Field:\\
Dielectric Response and Dynamics after Field Reversal}
\author{Kyohei Takae$^1$ and Akira Onuki$^2$}
%\email{takae@scphys.kyoto-u.ac.jp}
\affiliation{
$^1$Institute of Industrial Science, University of Tokyo,
4-6-1 Komaba, Meguro-ku, Tokyo 153-8505, Japan\\
$^2$Department of Physics, Kyoto University, Kyoto 606-8502, Japan}
\date{\today}

\begin{abstract} 
We study water between parallel metal walls  
under  applied electric field accounting for the image effect at $T=298$ K. 
The electric field due to the surface charges serves to  attract 
and orient nearby water molecules, while  it 
tends to a constant determined by the mean surface charge density 
away from the walls. We   find  Stern boundary layers 
with thickness about $5$ $\rm \AA$ 
 and a homogeneously polarized  bulk region.  
The molecules in the layers more sensitively 
respond to the applied field than in the bulk. 
As a result,  the potential drop  in the layers 
is larger than that in the bulk 
unless the cell length exceeds  10 nm. 
We also examine the  hydrogen bonds, which  tend to make 
small angles with respect to 
 the walls  in the layers even without applied field. 
%We also examine     the local electric field on each molecule 
%and its fluctuation  distribution. 
The  average  local field   
 considerably deviates  from the classical 
Lorentz field and the local field fluctuations are very large in the bulk.  
If we suppose  a nanometer-size sphere around each molecule, 
the local field contribution from its  exterior is nearly 
equal to that from   the continuum electrostatics and that from its interior 
yields the deviation from the classical Lorentz field. 
 As a  nonequilibrium problem, 
we investigate   the dynamics   after a reversal 
of  applied field, where  the relaxation is mostly 
caused by large-angle  rotational jumps after 1 ps  due to the presence of 
the hydrogen bond network. 
 The molecules undergoing these jumps  
themselves form  hydrogen-bonded   clusters  
heterogeneously distributed in space.
\end{abstract}

%\pacs{64.70.K-, 61.72.-y, 77.80.Jk, 77.65.-j}
%\keywords{}

\maketitle

%\tableofcontents
%\setcounter{tocdepth}{2}
%\pagebreak


%\setcounter{section}{1}  

\section{Introduction}

In  physics and chemistry, we need to  
 accurately estimate   the long-range electrostatic
interactions among charged and polar particles.
To this end, a large number of simulations have
been performed and the Ewald method 
is a famous technique for efficiently summing
these interactions using the Fourier transformation \cite{Allen}. 
It has been  used to investigate the bulk properties of 
charged and polar particles
under the periodic boundary condition 
in three dimensions (3D Ewald) \cite{Leeuw,Weis}.  
It has also been modified for film-like systems 
% between  non-polarizable and insulating walls 
%with the aid of two-dimensional Fourier transformation  
under the periodic boundary condition in 
the lateral directions (2D Ewald) \cite{Parry,Hey,Leew,Smith,Yeh1}. 
Several groups   \cite{Hautman,Perram,Klapp,Takae,Madden1,Yeh,Yeh1} have 
performed simulations of dipole 
systems in    electric field 
  between parallel metal  walls.  
However, not enough efforts have  been made on 
dynamics, where applied   electric field can be 
nonstationary.  Such nonequilibrium 
situations are ubiquitous and are  of great 
scientific and practical importance. 
Hence,  this paper  aims to give 
a general scheme of treating water 
under  electric field and 
investigate the dielectric  relaxation after field reversal.

Hautman {et al.}\cite{Hautman} developed  
a 3D  Ewald method 
assuming  parallel, smooth  metal walls, 
where the constant potential condition is satisfied at 
the metal walls ($z=0$ and $H$)  and the periodic boundary 
condition is imposed along the $x$ and $y$ axes. 
In this case, each charged particle 
in the cell  induces 
surface charges producing a potential equivalent to that
from an infinite number of image charges outside the cell. 
Perram and Ratner \cite{Perram} 
found  some relations on these image charges. 
In the same scheme,  Klapp\cite{Klapp}  treated dipoles interacting 
with the soft-core potential  to find wall-induced 
ordering. The present authors \cite{Takae}  extended 
this  3D  Ewald method  for charged and polar particles to  examine 
surface effects,  ionic crystals, 
  dipole chains, and local electric field. 
 In this paper, we use this method for water.

We also mention other   methods.  
Shelley and Patey \cite{Shelley} assumed 
  empty (vacuum) slabs   outside the  cell,
which the particles cannot  enter  
due  to  the repulsive wall potentials. 
If  the regions  $-d_{\rm em}<z<0$ and $H<z<H+d_{\rm em}$ 
are empty,  the  3D  Ewald method can be  used 
  with period $H+2 d_{\rm em}$  along  the $z$ axis.  
Also with   empty slabs,  
Yeh and  Berkowitz\cite{Yeh}  
 applied electric field  accounting for the local field 
from  net polarization. They found that   the computing time with 
this  3D Ewald  method was ten times faster than that with 
the 2D Ewald method.  With such  empty regions,  however,
charged or polar particles are effectively 
in contact with neutral, non-polarizable walls. 
Siepmann and Sprik \cite{Sprik} 
assumed  atomic particles   
forming a crystal  and   interacting   with 
water molecules via  a  model potential at the surface. 
They varied  charges of these atoms  continuously 
to  maintain the constant potential condition in metal. 
%Thus,  their electrodes model 
% incorporates the image  interaction. 
This  model  was  used  to study water and 
ions between  electrodes with the aid of   the  2D Ewald method 
 \cite{Reed,Madden1}. Petersen {\it et al.}\cite{Voth}  
proposed  an   efficient simulation method 
accounting for the primary image charges 
closest to  the boundary walls and 
 uniform (average) surface charge densities.

Surface charges  increase locally  as   charges or dipoles 
in the liquid region 
approach a  metal wall.   As a result,  water  molecules 
are  adsorbed  and  oriented near a metal wall\cite{Thiel,Hen,Sc}.
They form  a Stern surface  layer  \cite{Par,Beh} even without ions, 
where   the electric potential changes appreciably on a microscopic 
length. It also  follows  that the surface charges 
 exhibit significant  in-plane   fluctuations with a  correlation length 
$\xi_s$. We shall see  that the  
electric field due to these surface charges 
tends to be uniform in the bulk where  the distances from 
the walls much exceeds $\xi_s$.

We are not aware of previous  microscopic calculations 
to check  the  validity of the 	classical theory of dielectrics 
\cite{Fro,Kirk,Onsager}.  Hence, we   calculate the  
average and fluctuations of  the local electric field 
 ${\bi E}_k^{\rm loc}$ acting on each molecule $k$. 
%Supposing a sphere we divide  ${\bi E}_k^{\rm loc}$ 
% into the  contributions outside and inside the sphere. 
 In particular, we  consider  a nanometer-size  sphere 
surrounding each  molecule. In our simulation, 
 the  local field contribution  from the sphere interior  
consists of  an  average (a deviation from the classical  Lorentz field)   
and  large fluctuations ($\sim e/\sigma^2$ with $\sigma \sim 3~{\rm \AA}$), 
while that from the sphere exterior is 
obtained from  the continuum electrostatics with small fluctuations. 
 
We also present a first study  of  nonequilibrium water, 
where  the polarization relaxes  
after a reversal of applied electric field. 
In this  relaxation,  rotational jumps with 
large angle changes play a major role. These largely rotated molecules form 
clusters  causing   breakage and reorganization of the hydrogen 
bond network. In previous simulations on water, 
collective hydrogen-bond  dynamics was  
 studied  at $T \sim 300$ K \cite{Ohmine,Oh1} 
and marked dynamic heterogeneities were 
observed in translation and rotation  
in  supercooled   states  
\cite{Stanley,Mazza}.

The organization of this paper is as follows. In Sec. II,
we will reexamine  the Ewald scheme for water  between
metal walls. In Sec. III, we will calculate 
the dielectric response and the local electric field.  In Sec.IV, 
simulation results on field reversal will be presented. 
In Appendix A, we will give an expression for the 
local electric field on water molecules 
composed of three charge points. 
In Appendix B, we will devise  a microscopic expression 
for the polarization density ${\bi p}({\bi r})$, 
which is convenient for  theoretical study of water.

\section{Theoretical Background}

{\bf A. Water Model}. We use the TIP4P$/$2005 model\cite{Vega}, where 
each water molecule $k$ has three charge points 
${\bi r}_{k{\rm H}1}$,  ${\bi r}_{k{\rm H}2}$, and ${\bi r}_{k{\rm M}}$ 
with fixed partial charges  $q_{\rm H}$, $q_{\rm H}$, 
and $q_{\rm M}= -2q_{\rm H}$, 
respectively, where $q_{\rm H}=0.5564e$.
The point ${\bi r}_{k{\rm M}}$ is slightly 
shifted from the oxygen point ${\bi r}_{k{\rm O}}$ along 
${\bi n}_k$, where ${\bi n}_k$ is 
the unit vector along the bisector of 
the H-O-H triangle. Its  dipole moment  is given by 
\be 
 {\bi \mu}_k = q_{\rm H}( {\bi r}_{k{\rm H}1}+ 
{\bi r}_{k{\rm H}2}- 2{\bi r}_{k{\rm M}})= \mu_0{\bi n}_k,
\en 
where  $\mu_0=  2.305$D.  See Appendix A for more details. 
We adopt this fixed charge model because of its simplicity, though  
 the molecular polarizability is known to 
play a fundamental role in the properties of 
water\cite{Yu,polarizable}.

The total potential $U$  consists of  three parts as 
%\cite{Rossky} 
\bea 
U&=&U_{\rm m}+  \frac{1}{2}\sum_{k\neq \ell}u_{LJ}
(|{\bi  r}_{k{\rm O}}- {\bi r}_{\ell{\rm O}}|)\nonumber\\
&&+ \sum_{k} [u_{\rm w}(z_{k{\rm O}})+u_{\rm w}(H-z_{k{\rm O}})], 
\ena 
where $U_{\rm m}$ is the total electrostatic energy, 
$u_{\rm LJ}$ is the Lennard-Jones potential among the oxygen atoms,  
and $u_{\rm w}$ is the wall potential of the oxygen atoms: 
\bea 
u_{\rm LJ}(r)&=& 4\epsilon[ (\sigma/r)^{12} - 
(\sigma/r)^{6}],\\
u_{\rm w}(z) &=& C_9 (\sigma/z)^9-C_3 (\sigma/z)^3.
\ena  
We set $\epsilon=93.2k_B$, 
 $\sigma=3.1589{\rm \AA}$, 
$C_9=2\pi\epsilon/45$, and $C_3= 15C_9/2= \pi\epsilon/3$. 
Then,  the elementary charge is given by 
$e=23.82 (\epsilon\sigma)^{1/2}$.  
Due to $u_{\rm w}(z)$  distances of  any charge  positions 
 from the walls at $z=0$ and $H$ are larger than $1~{\rm \AA}$. 
The density and the orientation of 
water molecules near a wall sensitively depend 
on the form of $u_{\rm w}(z)$. 

In the literature,  extensive efforts have 
been made to examine surface states of water  
using various simulation methods 
\cite{Madden1,Reed,Sprik,Voth,Rossky,Thiel,Hen,Sc,Ber2}. 
We also remark that ab initio models  are needed to 
 accurately describe the surface 
potentials  on short length scales for water\cite{Ka}.

{\bf B. Electrostatic Energy and Image Charges}.
We consider a $L\times L\times H$ cell 
with metal plates at $z=0$ and $H$ using the 
periodic boundary condition along  the $x$ and $y$ axes. Its volume is 
$V=L^2H$. The walls at $z=0$ and $H$ 
are assumed to be  smooth and structureless for simplicity. 
We apply electric field under the fixed-potential condition.

Let   ${\bi r}_i=(x_i,y_i,z_i)$ and   ${\bi r}_j=(x_j,y_j,z_j)$  
 denote the 3N charge positions 
${\bi r}_{k{\rm H}1}$,  ${\bi r}_{k{\rm H}2}$, and ${\bi r}_{k{\rm M}}$ 
($1\le k\le N$). 
The electrostatic potential $\Phi({\bi r})$ outside 
the charge positions ${\bi r}\neq {\bi r}_i$ 
satisfies the metallic boundary condition, 
\be 
\Phi(x,y,0)=0, \quad 
\Phi(x,y,H)= -\Delta\Phi= -E_a H,
\en 
where  $\Delta\Phi$ 
is the applied potential difference and  $E_a=\Delta\Phi/H$ 
is the applied electric field. From eq 5, 
$\Phi({\bi r})$  is expressed in terms of image charges as 
%superposition of  Coulomb potentials and the applied field, 
\be 
\Phi =  \sum_{{\bi m}} {\sum_j}\bigg[
\frac{q_j}{|{\bi r}-{\bi r}_{j} + 
{\bi h}|} -   
\frac{q_j}{|{\bi r}-{\bar{\bi r}}_{j}  
 + {\bi h}|} \bigg]- E_a z,
\en   
where ${\bar{\bi r}}_j= (x_j, y_j, -z_j)$ and 
\be 
{\bi h}= (Lm_x,Lm_y,2Hm_z)
\en 
with  $m_x$, $m_y$, and $m_z$ being 
 integers ($0, \pm 1,\pm 2, ....)$. 
For  each real charge $q_j$ at ${\bi r}_j=(x_j,y_j,z_j)$ in the cell, 
we find  images with  the same  charge  $q_j$ 
at   $(x_j, y_j, z_j-2Hm_z)$ ($m_z= \pm 1, \cdots)$ 
and those with the opposite charge  $ -q_j $ at 
$(x_j, y_j, -z_j-2Hm_z)$ ($m_z=0, \pm 1, \cdots)$ outside the cell.  
Due to the summation over $m_z$  the metallic boundary condition 5 is satisfied.   

The electrostatic energy  
$U_{\rm m}$ at fixed $E_a$ is given by  
\cite{Hautman,Perram,Klapp,Takae}  
\be
U_{\rm m} =\frac{1}{2}
 \sum_{\bi m} \bigg[ {\sum_{ij}}'
\frac{q_iq_j}{|{\bi r}_{ij} + 
{\bi h}|}  
-   {\sum_{ij}}
\frac{q_iq_j}{|{\bar{\bi r}}_{ij}  
 + {\bi h}|} \bigg] -E_a M_z ,
\en  
where ${{\bi r}}_{ij}={\bi r}_i- 
{{\bi r}}_j$, ${\bar{\bi r}}_{ij}={\bi r}_i- 
{\bar{\bi r}}_j$, and  
\be 
M_z=  \sum_i q_i z_i= \sum_k \mu_{zk}
\en 
is the total polarization along the $z$ axis.
In  $\sum_{ij}'$ in eq 8, 
we exclude the self term with $j=i$ for  ${\bi h}=(0,0, 0)$. 
 For infinitesimal changes  
 ${\bi r}_i \to {\bi r}_i+ d{\bi r}_i$ and $E_a \to E_0+dE_a$, 
  the incremental change of $U_{\rm m}$ in eq 8 
is of the following differential form,    
\be 
d U_{\rm m} =  -\sum_i q_i {\bi E}_i \cdot d{\bi r}_i 
- M_z    dE_a,  
\en 
where ${\bi E}_i= - q_i^{-1}\nabla_i  U_{\rm m}$ 
is  the local electric field acting on charge $i$ at fixed $\Delta\Phi$. 
Hereafter, $\nabla_i=  \p /\p {\bi r}_i$. 
The local electric field ${\bi E}_k$ 
on molecule $k$ will be defined in Appendix A.

The first term in  eq 8 is periodic 
in 3D  with respect to  
$x_i \to x_i\pm L$, $y_i \to y_i\pm L$,  
and $z_i \to z_i\pm 2H$ due to the summation 
over ${\bi h}$.   Thus, it can  be calculated 
with the 3D  Ewald method. So far,  a few groups have 
performed simulations on the basis of $U_{\rm m}$ in eq 8. 
\cite{Hautman,Perram,Klapp,Takae}. 
In  the  Ewald method, the  Coulomb potential $ q_iq_j/r$  is divided
into  the long-range part $q_iq_j\psi_\ell(r)$ and 
the short-range part $q_iq_j\psi_s(r)$ with 
\be 
\psi_\ell(r) = {\rm erf}(\gamma r)/r, \quad 
\psi_s (r) =[1- {\rm erf}(\gamma r)]/r. 
\en 
where  ${\rm erf}(u)$ is 
the error function and $\gamma^{-1}$ 
represents  the  potential range of $\psi_s(r)$. 
It follows the Poisson equation 
 $-\nabla^2\psi_\ell(r)=4\pi \varphi_3(r)$ 
with $\varphi_3(r)=\varphi(x)\varphi(y)\varphi(z)$, where    
\be 
\varphi(z)=  (\gamma/\sqrt{\pi})  \exp( -\gamma^2 z^2) 
\en  
is the   1D 
Gaussian distribution with $\int dz\varphi(z)=1$.
 In this paper, we set  
$\gamma=0.85 /\sigma = 2.7$ $/$nm.

%1
\begin{figure}
\includegraphics[width=0.96\linewidth]{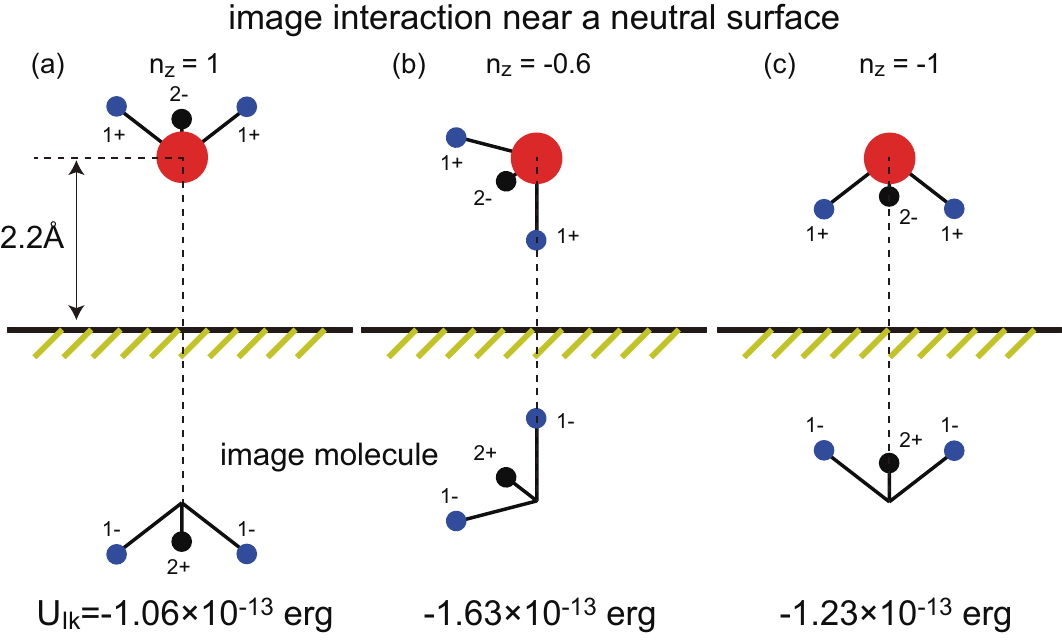}
\caption{ 
Three typical  configurations of a water molecule, whose 
 oxygen atom  is  separated by $2.2{\rm \AA}$ 
from a  metal wall without applied electric field. 
Image potential  $U_{{\rm I}k}$ in eq 13 is 
calculated to be (a) $-1.06$ ($n_z=1$), 
(b) $-1.63$ ($n_z=-0.6$),  and (c) $-1.23$ ($n_z=-1$) 
 in units of $10^{-13}$erg,  where $n_z=\cos\theta$ 
with  $\theta$ being the 
angle between the polarization direction and the $z$ axis. 
  H-down orientation (b)  
has the lowest electrostatic energy and most frequently 
appears close to the wall  without applied electric field. 
 }
\end{figure}

It is well-known that the image interaction 
grows when a charge or a dipole approaches 
a wall. For  a water molecule $k$ 
near  the bottom wall at $z=0$, 
the  image potential from the closest 
images  grows in  $U_{\rm m}$:  
\be 
U_{{\rm I}k}= -\sum_{\alpha \neq \beta }
\frac{q_\alpha q_\beta}{
2|{\bi r}_{k \alpha}- {\bar{\bi r}}_{k\beta}|},
\en   
where $ \alpha,\beta = {\rm H}1, {\rm H}2,{\rm M}$. 
In Fig.1,    typical molecular configurations 
are illustrated. At $z_{k{\rm O}}=2.2$${\rm \AA}$, 
the above  $U_{{\rm I}k}$ is lower for (b)   
than that for (a) by $5.7\times 10^{-14}$ erg ($=1.4k_BT$ for 
$T=298$ K).  Therefore, the image interaction favors 
the $H$ down configuration (b) near a metal wall, though 
 the molecular orientations near a wall 
are cooperative   due to the hydrogen bonding (see Fig.5).

{\bf C. Surface Charges}. 
The image charges are introduced as a mathematical 
convenience. The real charges are those in the cell and 
the surface charges (excess electrons)  on the metal walls. 
The latter  attract and orient dipoles  near the walls. 
Here, we examine the effects  of  the latter in detail. 
%near and 
% yield a nearly homogeneous electric field 
%far from the walls.  

The surface 
charge  densities are written as  $\sigma_0({\bi r}_\perp)$ at $z=0$ and 
$\sigma_H({\bi r}_\perp)$ at $z=H$, where ${\bi r}_\perp =(x,y)$. 
Since the charges in the cell are 
expelled from  the walls due to the wall potentials, 
$\sigma_0$ and $\sigma_H$ 
 are expressed in terms of $E_z(x, y, z)
=-\p \Phi/\p z$ as 
 \be 
\sigma_0= E_z(x, y, 0)/4\pi,\quad   
\sigma_H=-E_z(x, y,H)/4\pi.
\en 
We consider their lateral  mean surface charges,
\be 
{\bar{\sigma}}_0 = \frac{1}{L^2}\int dS 
{\sigma_0({\bi r}_\perp)}, \quad 
{\bar{\sigma}}_H= \frac{1}{L^2} \int dS 
{\sigma_H({\bi r}_\perp)}, 
\en 
where $dS=dxdy$ and $0<x,y<L$. From eq B4 in Appendix B, 
these mean values  exactly   satisfy \cite{Hautman,Takae}  
\be 
{\bar{\sigma}}_0= -{\bar{\sigma}}_H 
=  E_a/4\pi+  {M}_{z}/V .
\en  
We write the potential  from the 
 surface charges as $\Phi_s$ and   divide it  into two parts,  
\be 
\Phi_s({\bi r}) = -4\pi {\bar \sigma}_0 z+ \phi_s({\bi r}).
\en   
The  first term arises from 
the  mean surface charges. The  second contribution  
$\phi_s$  is due to   the surface charge  
deviations, which will be estimated in Sec.III.

In the   2D  Fourier series, we  set     
\bea 
&&\hspace{-12mm}\Delta \sigma_0({\bi r}_\perp)
= \sigma_0({\bi r}_\perp)- {\bar{\sigma}}_0=    
\sum_{{\bi k}\neq{\bi 0}} \sigma_{0{\bi k}}
 e^{{{\rm i}{\bi k}\cdot{\bi r}_\perp}},\nonumber\\
&&\hspace{-12mm}\Delta 
\sigma_H ({\bi r}_\perp)=\sigma_H ({\bi r}_\perp)-  {\bar{\sigma}}_H=  
\sum_{{\bi k}\neq{\bi 0}}
\sigma_{H{\bi k}}e^{{\rm i}{\bi k}\cdot{\bi r}_\perp}, 
\ena  
where  ${\bi k}=(2\pi /L)(n_x,  n_y)\neq (0,0)$ with $n_x$ 
and $n_y$ being  integers. 
%where $n=N/V$ and ${\bar \mu}_{z}= \sum_i \mu_{iz}/N$ is the average.  
%From eq 2.5), 
The Fourier components   are calculated as\cite{Takae}    
%In terms of the charge positions  ${\bi r}_j$  we have 
\bea 
&&\hspace{-4mm} \sigma_{0{\bi k}}= 
\frac{ -1}{L^2} \sum_j q_j  
\frac{\sinh(kH-kz_j)}{\sinh(kH)}  e^{-{\rm i}{\bi k}\cdot 
{\bi r}_j},   
\nonumber\\
&&\hspace{-4mm} \sigma_{H{\bi k}}=   
\frac{-1}{L^2} \sum_j q_j 
\frac{\sinh(kz_j)}{\sinh(kH)}  e^{-{\rm i}{\bi k}\cdot {\bi r}_j},   
\ena 
where   $k= |{\bi k}|\neq 0$  and the summation over $m_z$ in eq 7 
has been performed to give the hyperbolic sine functions. 
The potential deviation $\phi_s$ is expressed as 
\bea 
&&\hspace{-5mm}\phi_s
= \sum_{{\bi m}_\perp}
\int dS'  \bigg[\frac{\Delta\sigma_0({{\bi r}_\perp}')
}{|{\bi r}-{\bi r}'+L{\bi m}_\perp|} +\frac{\Delta\sigma_H({{\bi r}_\perp}')
}{|{\bi r}-{\bi r}''+L{\bi m}_\perp|} \bigg]\nonumber\\
&&\hspace{0mm}= \sum_{{\bi k}\neq {\bi 0}}
 \frac{2\pi }{k} \bigg 
[ \sigma_{0{\bi k}}e^{-kz}+ \sigma_{H{\bi k}}e^{k(z-H)}\bigg] 
 e^{{\rm i}{\bi k}\cdot{\bi r}_\perp}.
\ena
In the first line, $dS'=dx'dy'$, ${{\bi r}_\perp}'=(x',y') $, 
 ${\bi r}'= (x',y',0)$,  ${\bi r}''= 
(x',y',H)$, and  ${\bi m}_\perp= (m_x,m_y,0)$ 
with $m_x$ and $m_y$ being integers. 
 The second line is the 2D Fourier expansion of the first line, 
where we use the 2D integral $ \int dS 
e^{{\rm i}{\bi k}\cdot{\bi r}_\perp}/r=2\pi e^{-kz}/k$.

The  total potential $\Phi({\bi r})$ (${\bi r}\neq {\bi r}_j$) 
 in eq 6  arises from the charges in the cell 
and those on the walls as 
\be 
\Phi =   \sum_{ {\bi m}_\perp} 
 {\sum_{j}}'  
\frac{q_j}{|{\bi r}- {\bi r}_{ j} +  L{\bi m}_\perp|} +\phi_s({\bi r})
-4\pi {\bar \sigma}_0 z ,
\en 
which also follows if we  substitute 
 eq 19 into the second line of eq 20.  The summation over 
 ${\bi m}_\perp$ in the first term ensures the lateral periodicity. 
From eq 9,  $\phi_s({\bi r})$ is also 
written in terms of the image potentials as 
\bea 
\phi_s &=& 
 \sum_{{\bi m}} {\sum_j}\bigg[
\frac{q_j(1-\delta_{m_z 0})}{|{\bi r}-{\bi r}_{j} + 
{\bi h}|} -    \frac{q_j}{|{\bi r}-{\bar{\bi r}}_{j}
 + {\bi h}|} \bigg] \nonumber\\  
&&+{4\pi M_z z}/{V} ,
\ena   
where $\delta_{mn}$ is   the Kronecker delta and 
$\bi h$ is expressed as eq 7. 
The first term  in eq 22  is the sum of  
the image potentials.  Far from the walls, 
we shall see that  it  is mostly 
 canceled by  the second term ($\propto M_z$), 
leading to small $\phi_s$ (see Sec.III).

Using eq 22, we rewrite  
the electrostatic  energy $U_{\rm m}$ in eq 8 
using  $\phi_s$ as  
\bea 
U_{\rm m}  &=&  \frac{1}{2} \sum_{ {\bi m}_\perp} 
 {\sum_{i,j}}'  
\frac{q_i q_j}{|{\bi r}_i- {\bi r}_{j} +  L{\bi m}_\perp|}
+ \sum_j  \frac{q_j}{2} \phi_s({\bi r}_j) 
\nonumber\\
&& -2\pi M_z^2/V-  {E_a} {M}_z, 
\ena 
The third  term ($\propto M_z^2$)   in eq 23  
 is a   mean-field contribution, which 
is large in applied field. 
Using  the relation $\nabla_i \sum_j 
{q_j} \phi_s({\bi r}_j)= -2q_i {\bi E}_s({\bi r}_i)$,   
the local field ${\bf E}_i$ in eq 10 is divided  as  
\be 
{\bi E}_i={\bi E}_i^d   + {\bi E}_s ({\bi r}_i) + 
4\pi{\bar \sigma}_0 {\bi e}_z.
\en 
The ${\bi E}_i^d$ arises from  the 
charges in the cell   and ${\bi E}_s({\bi r})$  
from  the surface charge deviations so that    
\bea 
&&\hspace{-3mm} 
{\bi E}_i^d =  \sum_{ {\bi m}_\perp} 
 {\sum_{j}}'  { q_j}{\bi g}({{\bi r}_i- {\bi r}_{j} +  L{\bi m}_\perp}) ,  
\\
&&\hspace{-3mm} {\bi E}_s({\bi r})= -\nabla \phi_s ({\bi r}).
\ena
where 
%$\nabla_i= \p/\p {\bi r}_i$ and 
${\bi g}({\bi r}) = -\nabla r^{-1}=r^{-3}{\bi r}$. 
The third term in eq 24 arises from 
 the mean surface charge,  
%related to $M_z/V$ by eq 2.14), 
where  
 ${\bi e}_z$ is  the unit vector along the $z$ axis. 
 In Sec.IIIE, we will further divide 
${\bi E}_i^d $ into long-range and short-range parts.

Yeh and  Berkowitz\cite{Yeh}  used the 3D Ewald method 
with empty slabs  
under  applied electric field. 
They replaced  $E_a$  by $E_a+ 4\pi M_z/V$ in 
the equations of motion, so their method  is 
 justified by  eqs 23  and 24  provided that 
 $\phi_s$ is  negligible far from the walls.

\section{Equilibrium States under Electric Field}
We performed MD simulation using $U_{\rm m}$ in eq 8 
with  the 3D Ewald method. 
The molecule number is $N=2400$ 
and the cell dimensions are $L=41.5$${\rm \AA}$ and 
$H=44.7$$\rm \AA$ with  volume  
$V=L^2H \cong 77$nm$^3$.  
%The mass density is $1.01$ g$/$cm$^3$. 
  The temperature is fixed at $T=298$ K in the $NVT$ ensemble with a 
Nos$\acute{\rm{e}}$-Hoover thermostat.    
See  the beginning of  Sec.IVA  for  remarks on  the $NVE$ simulation.   
In this section, the symbol  $\av{\cdots}$ denotes  
 the time   average over   6 ns,  
which is taken as  the equilibrium  average.

%2
\begin{figure}
\includegraphics[width=0.96\linewidth]{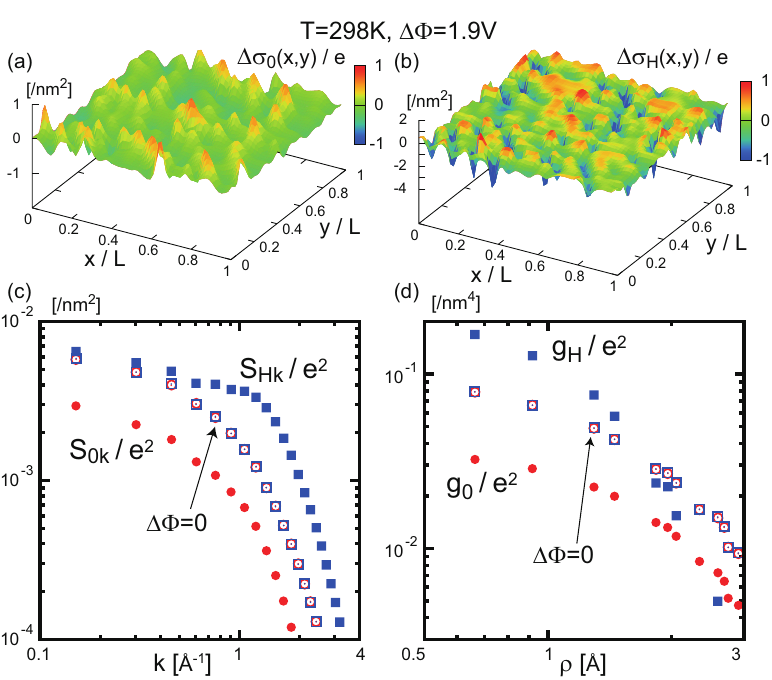}
\caption{ 
 Snapshots of surface charge deviations (a) $\Delta\sigma_0(x,y)$  
at $z=0$  and (b) $\Delta\sigma_H(x,y)$ at $z=H$  divided by $e$ at $T=298$K, 
 where $\Delta \Phi=1.9$V ($E_a=0.42$ V$/$nm).  
(c) Surface-charge structure factors $S_{0k}$   and 
$S_{Hk}$   in eq 27 vs $k$ and (d) surface-charge spatial  
correlation functions 
$g_{0}(\rho)$   and $g_{H}(\rho)$  in eq 28  
vs $\rho=(x^2+y^2)^{1/2}$, where 
 $\Delta \Phi=0$ and $1.9$ V.  
These quantities are divided by $e^2$. }
\end{figure}

{\bf A. Effects of Surface Charges}.  
In Fig.2, we present snapshots of (a) 
the surface charge deviations $\Delta\sigma_0(x,y)$  at $z=0$ 
 and (b) $\Delta\sigma_H(x,y)$  at $z=H$ with 
$\Delta\Phi=1.9$ V or $E_a= 0.42$ V$/$nm,   where 
the bottom (top) wall 
is positively (negatively) charged with $\bar{\sigma}_0/e=0.44/$nm$^2$. 
Here,   the fluctuation amplitude 
of  $\Delta\sigma_H$ is larger than that  of $\Delta\sigma_0$,  
 because the protons  can be closer to  
the  top wall  than the  oxygen atoms to  the bottom  wall.  
%(See Fig.12 for examples of the molecular 
%configurations near the bottom.)
In (c), we display  the 2D structure factors $S_{0  k}$ 
and  $S_{H  k}$ for  the thermal fluctuations of 
  the surface charges  defined     by  
\be 
S_{0  k}= {L^2} \av{|\sigma_{0{\bi k}}|^2}, \quad 
 S_{Hk}= {L^2} \av{|\sigma_{H{\bi k}}|^2} .
\en 
which  depend only on $k=|{\bi k}|$ for $kL\gg 1$.
Here, $S_{H  k}$  is considerably larger than $S_{0  k}$ 
for $\Delta\Phi=1.9$ V, while they coincide for $\Delta\Phi=0$. 
Setting $S_{\lambda 0}/S_{\lambda k}= 1+ \xi_{s\lambda}^2k^2+ \cdots$ for 
small $k$, we determine  the correlation lengths $\xi_{s0}$ and 
 $\xi_{sH}$. Then,     $\xi_{s0}=1.95~{\rm \AA}$ 
and $ \xi_{sH}=0.95~{\rm \AA}\sim 0.5 \xi_{s0}$  
for $\Delta\Phi=1.9$ V, while 
$\xi_{s0}=\xi_{sH}=  
1.6~{\rm \AA}$ for $\Delta\Phi=0$.   
In (d), we present 
the corresponding  2D pair correlation functions $g_0(\rho )$ 
and $g_H(\rho )$  expressed as  
\be
g_\lambda(\rho )=\av{\Delta\sigma_\lambda({\bi 0})\Delta 
\sigma_\lambda( {\bi r}_\perp)}
 = \frac{1}{L^2} \sum_{{\bi k}\neq {\bi 0}}
S_{\lambda k}e^{{\rm i}{\bi k}\cdot{\bi r}_\perp}, 
\en
where $\lambda=0, H$, and  ${\bi r}_\perp=(x,y)$,  
 These functions depend only  on $\rho=|{\bi r}_\perp|=
({x^2+y^2})^{1/2}$  for $\rho\ll L$. 
Here, $g_0(0)/e^2 = \av{|\Delta\sigma_0|^2}/e^2 
= 0.037  /$nm$^4$ and $g_H(0)/e^2 = \av{|\Delta\sigma_H|^2}/e^2 
= 0.224 /$nm$^4$ for $\Delta\Phi=1.9$ V  in accord with 
the amplitude difference in (a) and (b), while 
these quantities are about $0.095/$nm$^4$ for $\Delta\Phi=0$.

%3
\begin{figure}
\includegraphics[width=0.96\linewidth]{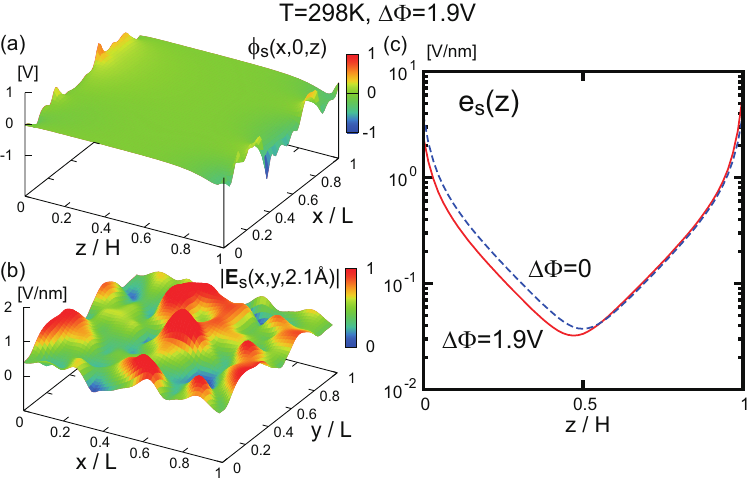}
\caption{Snapshots of (a) electric potential $\phi_s(x,0,z)$ 
due to the surface charge deviations at $y=0$ in the $zx$ plane 
and (b)  {$|{\bi E_s(x,y,z)}|=|\nabla\phi_s|$}  at $z=2.1{\rm \AA}$  
in the $xy$ plane, where 
 $\Delta \Phi=1.9$V ($E_a=0.42$ V$/$nm) 
and $T=298$K as in Fig.2. 
(c) Electric-field fluctuation  amplitude 
$e_s(z)$ in  eq 29.
 }
\end{figure}

We are  also interested in the electric field created by  
 the surface charge deviations. 
 In Fig.3, we  show   examples of cross-sectional 
snapshots of (a) $\phi_s(x,0,z)$  in the $zx$ plane  and 
(b) $|{\bi E}_s(x,y,2.1~{\rm \AA})|$  in the $xy$ plane. 
Here, $\phi_s$ tends to zero far from the walls, 
while $\phi_s$ and  ${\bi E}_s$  fluctuates near the walls. 
Using  the second line of eq 20, 
we  introduce   the  fluctuation  amplitude $e_s(z)$ of 
${\bi E}_s({\bi r}) =-\nabla\phi_s({\bi r})$  by  
\be
%\hspace{-0.2mm} 
e_s(z)^2=
\av{|{\bi E}_s|^2}
= 
\frac{8\pi^2  }{L^2} 
\sum_{{\bi k}\neq {\bi 0}}
\sum_{\lambda=0,H} S_{\lambda  k} e^{-2k|z-\lambda|} . 
\en    
where  $e_s(z)$  depends only on $z$ due to the averages 
in time and in the $xy$ plane.  
In (c), we display $e_s(z)$ for $\Delta\Phi=0$ 
and 1.9 V, which  becomes very small away from the walls. 
In Appendix C, we will examine the behavior of 
$e_s(z)$ away from the walls in more detail.

{\bf B. Average 1D Polarization and  Potential}. 
In Appendix B, we will give  the microscopic expression for 
the polarization density 
${\bi p}({\bi r})$ for water molecules. In our 1D geometry,  the  average 
polarization along the $z$ axis is equal to 
the   average in the $xy$ plane:   
\be
{P}(z) =  \av{p_z({\bi r})}= \frac{-1}{L^2}
\sum_j \av{ q_j \theta(z- z_{j})}, 
\en 
where $\theta(u)$ is  the step function 
being 0 for $u\le 0$ and 1 for $u>0$. 
We may then define  
 the average Poisson electric  potential $\Psi(z)$ 
 and field ${\cal E}(z)=- d\Psi/dz$ as \cite{Hautman,Yeh,Madden1} 
\bea 
\Psi(z)&=&- 4\pi \av{{\bar{\sigma}}_0} z + 4\pi \int_0^z dz' P(z'),\\
{\cal E}(z)&=&
 4\pi \av{{\bar{\sigma}}_0} -4\pi  P(z). 
\ena  
where   $\av{{\bar{\sigma}}_0}$ is 
 the average of ${{\bar{\sigma}}_0}$.  
In our case, 
the temporal fluctuations of ${\bar{\sigma}}_0$  
are very small and we need not distinguish 
between  ${{\bar{\sigma}}_0}$ and $\av{{\bar{\sigma}}_0}$. 
 From eqs 30 and 31, 
 $\Psi(z)$ satisfies $\Psi(0)=0$, $\Psi(H)= - E_a H$, and 
\be 
 \frac{d^2\Psi(z)}{dx^2} = {4\pi} \frac{d  P(z)}{dz}
= -\frac{4\pi}{L^2} \sum_j \av{q_j \delta (z-z_j)}. 
\en 
Taking the average of eq 16 gives    
\be 
\av{{\bar{\sigma}}_0}=-\av{{\bar{\sigma}}_H} =E_a/4\pi+ \int_0^H dz P(z)/H,  
\en 
 where $\int_0^H dz P(z)= \av{M_z}/L^2$.
The    $P(z)$, $\Psi(z)$, and ${\cal E}(z)$ correspond to the 
polarization, electric potential, and electric field 
in the continuum electrostatics. The  effective 
 dielectric constant of a film may be  defined in terms of 
$\av{{\bar\sigma}_0}$ or $\av{M_z}$ as  \cite{Hautman,Madden1} 
\be
\ve_{\rm eff}= 4\pi\av{{\bar\sigma}_0}/E_a
=1+ {4\pi}\av{M_z}/VE_a.
\en 
In  addition, $U_{\rm m}$ in eq 8 or eq 23 
yields  $\av{M_z}= \av{M_z^2}_0 E_a/k_BT$ in the limit of small  $E_a$, 
where $\av{\cdots}_0$ is the equilibrium average with $\Delta\Phi=0$.
 Thus, $\av{M_z}_0=0$. 
The linear response expression for $\ve_{\rm eff}$ is given by 
\cite{Neu,St,Vega,polarizable} 
\be
\lim_{\Delta\Phi \to 0}\ve_{\rm eff}= 1+ {4\pi}\av{M_z^2}_0/Vk_BT.
\en

%4 
\begin{figure}
\includegraphics[width=0.96\linewidth]{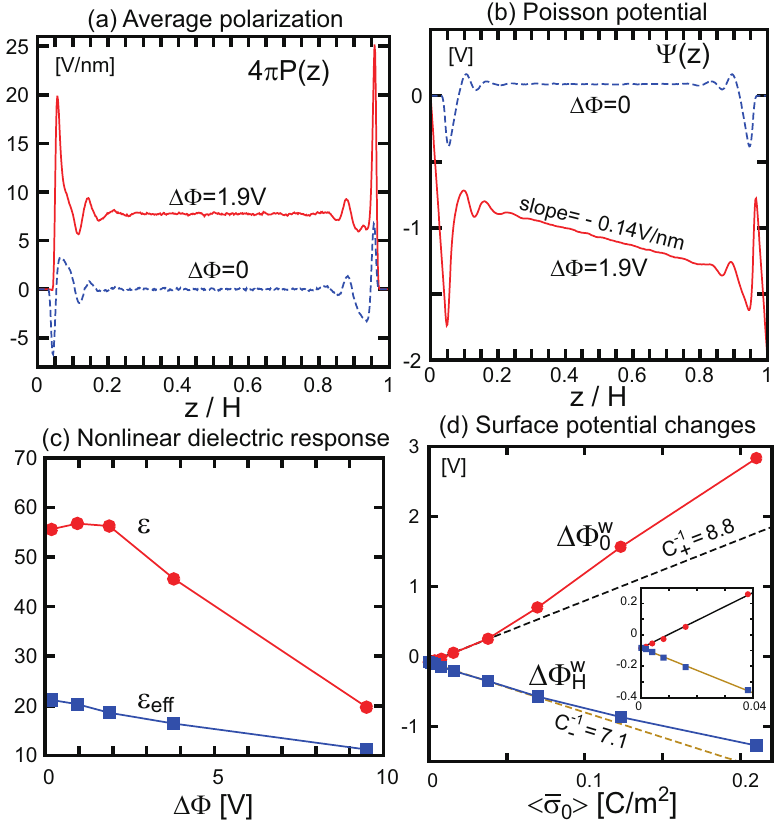}
\caption{ Average  1D  profiles  
at $T=298$ K.  (a) $4\pi P(z)$ in eq 30  and 
(b) $\Psi(z)$ in eq 31  for $\Delta\Phi=0$ and 1.9 V. 
(c) $\ve_{\rm eff}$ in eq 35 and $\ve$ in eq 38 vs 
$\Delta\Phi$.  (d) $\Delta \Phi_{0}^{\rm w}$ and  $\Delta \Phi_{H}^{\rm w}$ 
in eq 40 vs $\av{{\bar \sigma}_0}$. In the inset,  small-field behavior 
is expanded, indicating  the zero-field limit $\Phi_{00}^{\rm w}\cong 
-0.09$ V  in eq 44.  }
\end{figure}

In Fig.4,  (a) $4\pi P(z)$  and  
(b) $\Psi(z)$ are displayed  for $\Delta\Phi=0$ and 1.9 V. 
Here, there appear  Stern boundary layers \cite{Par,Beh} 
 near the walls. Its  thickness is given by $d= 4.7{\rm \AA}$ if 
it is defined as the first maximum  distance of $\Psi(z)$ for $\Delta\Phi=0$. 
Even for  $\Delta\Phi=1.9$ V, the layer  thickness 
 is of this order, while  $\Psi(z)$ becomes   oscillatory near the walls. 
Outside the layers ($z>d$ and $H-z>d$),  
   a  homogeneous (bulk) state    
is realized  with  thickness $H-2d(=3.5$ nm here)
\cite{Hautman,Madden1,Yeh,Takae},  where 
$P(z)$ and ${\cal E}(z)$ are   constant as   
\be 
P(z)\cong P_b, \quad {\cal E}(z)\cong  E_b= 
4\pi (\av{{\bar{\sigma}}_0} -  P_b).
\en 
The  $P_b$ and $E_b$ are the bulk polarization 
and electric field, respectively. 
The bulk  dielectric constant is given by\cite{Yeh} 
\be
 \ve =4\pi\av{{\bar\sigma}_0}/E_b= 1+ 4\pi P_b/E_b. 
\en  
In Fig.4(c), we show $\ve_{\rm eff}$ and   
$\ve$ vs $\Delta\Phi$, where they exhibit 
 considerable nonlinear  behavior with increasing $\Delta\Phi$. 
We find $\ve\sim 60$ for small $\Delta\Phi$ 
as in the previous simulations \cite{Yeh,Yeh1,St,Madden1}.
However, as $\Delta\Phi\to 0$, $E_b$ becomes very small and 
 $\ve$ cannot be determined precisely from eq 38. 
As regards $\ve_{\rm eff}$, the right hand side of 
the linear response formula (36) is calculated to be  21.4, 
%26, This is incorrect 
which  coincides  with   $\ve_{\rm eff}$ from eq 35  at small 
$\Delta\Phi$  in (c).
%In   the nonlinear regime,  the dipole moments  ${\bi \mu}_k= 
%\mu_0 {\bi n}_k$ largely align  along the $z$ axis. 
In the three regions  
$z<d$ (bottom), $0.3<z/H<0.7$ (bulk), and $H-z<d$ (top), 
the average of the $z$ component ${n_{zk}}$ 
is given by  $0.02,0$, and $-0.02$  for $\Delta\Phi=0$ 
and by $0.35, 0.27$, and $0.33$  for $\Delta\Phi=1.9$ V, respectively. 
 The dipoles are more aligned  in the Stern layers 
than in the bulk with small asymmetry  between bottom and top. 
The alignment     tends to  saturate  with increasing $\Delta\Phi$. 
%   bottom  bulk  top
%0V    0.02    0    -0.02
%1.9V  0.36    0.27  0.33
%3.8V  0.53    0.48  0.53
The average density in the bulk region is 
%$1.071\sigma^{-3}=
$33.97$ nm$^{-3}=1.016$ g$/$cm$^3$ for 
$\Delta\Phi=0$ and 
%$1.058\sigma^{-3}
$33.56$ nm$^{-3}=1.004$ g$/$cm$^3$ for 
$\Delta\Phi=1.9$ V.

In the bulk,  $\Psi(z)$  behaves linearly as    
\be 
\Psi(z) \cong  -E_b z -  A_0,  
\en 
where  $A_0$ is a constant. We determine 
  the excess  potential changes in the Stern    layers  
at $z=0$ and $H$  as 
\cite{Hautman,Madden1} 
\bea 
&&  \Delta \Phi_{0}^{\rm w}=\Psi(0)-\Psi(d) -E_b d,\nonumber\\ 
&&  \Delta \Phi_{H}^{\rm w}= \Psi(H)-\Psi(H-d)+E_b d.
\ena 
Then, $A_0=  \Delta \Phi_{0}^{\rm w}$ in eq 39. 
Use of  $P(z)$ gives  
\be 
 {\Delta \Phi_{\lambda}^{\rm w}}=
{4\pi}\int_\lambda^{|\lambda-d|} dz [P_b-P(z)],
\en 
where $\lambda=0,H$. Since $E_aH= \Phi_0-\Phi_H= 
 \Delta \Phi_{0}^{\rm w}- \Delta \Phi_{H}^{\rm w}
+ E_b H$, we  define a surface  electric length by  
\be 
\ell_{\rm w}=   (\Delta\Phi_0^{\rm w}-\Delta\Phi_H^{\rm w})/E_b. 
\en 
The potential drop in the bottom and top layers is given by 
$\Delta\Phi \ell_{\rm w}/(\ell_{\rm w}+H)$. 
In terms of the ratio $\ell_{\rm w}/H$, the bulk quantities 
 $\ve$,  $E_b$, and  $P_b$ can be  related to  the corresponding 
film quantities $\ve_{\rm eff}$,  $E_a$, and  $\av{M_z}/V$  as  
\bea 
&&\ve/\ve_{\rm eff}=E_a/E_b = 1+\ell_{\rm w}/H, \\
&& P_b= \av{M_z}/V +  E_b\ell_{\rm w} /4\pi H.
\ena 

In  Fig.4(b),  there arises 
 a surface potential change even for $\Delta\Phi=0$. 
We define   the zero-field surface potential drop as    
\be 
\Phi_{00}^{\rm w}=  \lim_{\Delta\Phi\to 0} 
\Delta\Psi_0^{\rm w} =   \lim_{\Delta\Phi\to 0} \Delta \Psi_H^{\rm w}. 
\en  
For  $\Delta\Phi\ge 0$ or for $\av{{\bar\sigma}_0}\ge 0$, 
we relate $\Delta \Phi_{0}^{\rm w}$ and $\Delta \Phi_{H}^{\rm w}$ 
 to $\av{{\bar\sigma}_0}$ and $\av{{\bar\sigma}_H}= - \av{{\bar\sigma}_0}$ 
as   
\bea 
&& \Delta \Phi_{0}^{\rm w}=\Phi_{00}^{\rm w} +C_+^{-1} \av{{\bar\sigma}_0}, \nonumber\\ 
&& \Delta \Phi_{H}^{\rm w}= \Phi_{00}^{\rm w} +
C_-^{-1} \av{{\bar\sigma}_H},
\ena 
where $C_+$ and  $C_-$ are the surface capacitance    
\cite{Par,Beh} for positive and negative surface charges, respectively. 
From eqs 42 and 46 with the aid of eq 38, we  obtain 
\be 
\ell_{\rm w}= (C_+^{-1}+C_-^{-1})\ve/4\pi .
\en

\begin{table}
\caption{{\bf Data  for water in equilibrium under electric field.} 
For $\Delta\Phi=0.19$ V 
and 1.9 V at $T=298$ K,  listed are 
$E_a$ (V$/$nm), $E_b$ (V$/$nm), $4\pi P_b$ (V$/$nm),
$4\pi\av{{\bar\sigma}_0}$ (V$/$nm),  $\ve_{\rm eff}$, 
$\ve$, $\Delta\Phi_0^{\rm w}$ (V), and
$\Delta\Phi_H^{\rm w}$ (V).  These are defined in 
eqs 35-40, while  ${\gamma_{\rm loc}}$ is the 
Lorentz factor in bulk 
%for the average  local electric field 
in eq 49. 
 Instead of $4\pi\av{{\bar\sigma}_0}$ in V$/$nm, 
we also have  
$\av{{\bar\sigma_0}}/e=0.05 $ and $ 0.44$, respectively,   
in  nm$^{-2}$.  
  }
\begin{tabular}{|c|c||c|c|c|c|c|c|c|c|} 
\hline
$\Delta\Phi$
  &$E_a$&$E_b$&$4\pi P_b$ &$4\pi\av{{\bar\sigma_0}}$&$\ve_{\rm eff}$ &$\ve$
&$\Delta\Phi_0^{\rm w}$&$\Delta\Phi_H^{\rm w}$
&${\gamma_{\rm loc}}$\\
%&1& 2& 3 \\
%\hline
%0.19 &0.042 &0.017 &0.88 &0.05&21&53&-0.029&-0.14&0.64 \\
%\hline
%1.9&0.42& 0.14&7.7 &0.44&19&56 &0.71&-0.56&0.64\\
\hline
0.19 &0.042 &0.016 &0.88 &0.90 &21 &56 &-0.026 &-0.14 &0.58 \\
\hline
1.9 &0.42 &0.14 &7.7 &7.9 &19 &56 &0.70 &-0.57 &0.59 \\
\hline
\end{tabular}
\end{table}

In Table 1, we give examples of numerical values of 
   $E_b$, $4\pi P_b$, $4\pi \av{{\bar\sigma}_0}$, 
$\ve_{\rm eff}$,  $\ve$,  $\Delta\Phi_0^{\rm w}$, 
$\Delta\Phi_H^{\rm w}$, and ${\gamma_{\rm loc}}$ 
(see eq 49) for ${\gamma_{\rm loc}}$), setting  
$\Delta\Phi=0.19$ and 1.9 V 
($E_a=0.042$ and 0.42 V$/$nm)  at $298$ K. 
In Fig.4(d),we  have 
 $\Phi_{00}^{\rm w}=-0.46(\epsilon /\sigma)^{1/2}=-0.019e/\sigma=
-0.09 {\rm V}$. 
We also have the surface capacitance   
$C_+=1.0/{\rm nm}$ and   $C_-=1.4/{\rm nm}$ for 
 $\av{{\bar \sigma}_0 } <0.04$ C$/$m$^2= 0.24 e/$nm$^2$  
or for $\Delta\Phi<1$ V, 
but they exhibit considerable nonlinear behavior 
for larger $\av{{\bar \sigma}_0 }$.   
In the usual units, they are written  as 
$C_+= 11 \mu {\rm F} /{\rm cm}^2$,
and $ 
C_-=14 \mu {\rm F}/{\rm cm}^2$  for 
small $\av{{\bar \sigma}_0 }$. 
%where $C_+^{-1}=4.7 \sigma=  14.1$ and $C_-^{-1}=2.1= 6.4$
%(14.1+ 6.4)*60/4\pi  , \sigma=3.1589
 Thus, from eq 47, 
we find 
$\ell_{\rm w}=8$ ${\rm nm}\sim 2H$ for small 
$\av{{\bar \sigma}_0 }$, which is very long.  As a result, we have 
$E_a/E_b= \ve/\ve_{\rm eff}\sim 3$ 
for our small system. 
%However,    $C_+$ and  $C_-$ behave  
%nonlinearly  for large $\Delta\Phi$. 
In contrast, in our recent 
paper\cite{Takae}, we applied electric field 
to point dipoles interacting with the soft-core potential 
to obtain $\ell_{\rm w}\sim 2\sigma \sim  0.1H$ 
and $\ve_{\rm eff}\cong \ve$.

Hautman {\it et al.}\cite{Hautman} obtained 
% $\Phi_{00}^{\rm w} = - 0.58$ V, 
$C_+= 12.4 $ $\mu F /$cm$^2$ 
 and $C_-= 10.5 $ $\mu F /$cm$^2$  using the SPC model, 
where    $\ve_{\rm eff}$ in  eq 35 
was  $12$ in one example in accord with eq 43. 
Yeh and Berkowitz \cite{Yeh,Yeh1}  obtained  $\ve\sim 60$ for small 
$\Delta\Phi$.   Willard {\it et al.} \cite{Madden1} 
obtained    $\ve \sim 75.07$, 61.50, and 57.30 
for  $\Delta\Phi=$0.27, 1.36 and 2.72 V, respectively, 
together with $C_+ = 5.20$ $\mu F$$/$cm$^2$  
and $C_- = 8.39$ $\mu F$$/$cm$^2$   
using the SPC$/$E model.  
%where  the maximum was  75 at small $\Delta\Phi$. 
We  may   estimate  $\ell_{\rm w}$ in eq 47 
 using  their data \cite{Hautman,Madden1} to 
find $\ell_{\rm w} \sim 10$ ${\rm nm}$ as in our case. 
These numerical values of the surface capacitance   
are considerably smaller than the experimental values  
for electrolytes in contact with  a metal surface 
\cite{Hautman,Madden1,Par,Beh}.

%5
\begin{figure}
\includegraphics[width=0.96\linewidth]{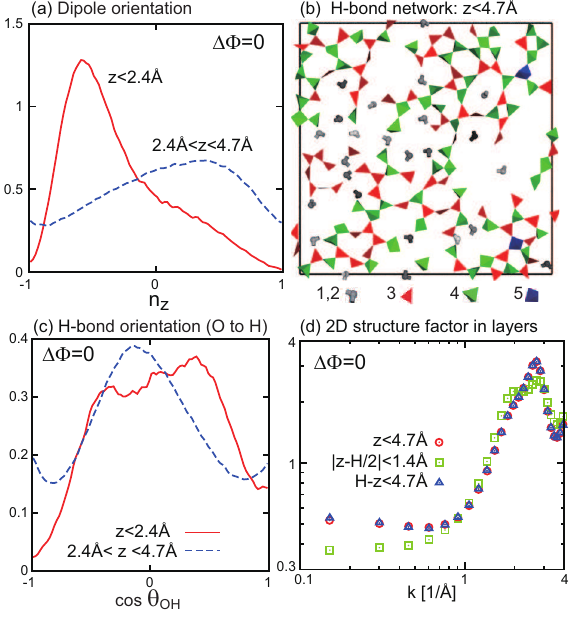}
\caption{   
(a) Distributions of  
the $z$ component of the polarization  direction ${n}_{zk}$ 
in the first  ($z<2.4$ ${\rm \AA}$) 
and  second ($2.4$ ${\rm \AA}<z<4.7$ ${\rm \AA}$) layers. 
(b)   Snapshot of 167  molecules in the bottom Stern layer 
with  intermolecular hydrogen bonds, whose numbers 
are $1$, 2, 3, 4, and 5 as indicated below the panel. 
(c) Distributions of $\cos\theta_{k{\rm OH}}$ 
 indicating planar alignment, where 
 $\theta_{k{\rm OH}}$ is the intermolecular  
hydrogen bond angle with respect to the $z$ axis 
(from O to H).
 (d) 2D structure factors 
of the molecular positions $(x_{k{\rm G}},y_{k{\rm G}})$ 
in three layer regions 
$z<4.7$ $\rm \AA$, $|z-H/2|<1.4$  $\rm \AA$, 
and  $H-z<4.7$  $\rm \AA$, indicating planar 
mesoscopic  fluctuations near the walls. 
 } 
\end{figure}

For $\Delta\Phi=0$, we may    treat  the Stern layer  as a bilayer 
composed of two parts,    $z<d_1=$2.4 ${\rm \AA}$ 
 and $d_1<z<d=$4.7 ${\rm \AA}$,  at the bottom. 
 Here,  $\Psi(z)$ in  eq 31 takes a minimum at 
$z=d_1$ and a maximum at $z=d$ in Fig.4(b).
 In the first and second layers, 
the  average areal density of water molecules 
and the average of  $n_{zk}=\mu_{zk}/\mu_0$ are given by 
$(1.6$ ${\rm nm}^{-2}$, -0.26)
and ($8.2$ ${\rm nm}^{-2}$, 0.073), respectively.  
See  Fig.5(a), where displayed are  the normalized distributions 
of  $n_{zk} $ in the two layers. 
The polarization is downward in the first layer 
and upward in the second. We remark that 
the zero-field potential drop is expressed as  
 $\Phi_{00}^{\rm w}= -4\pi\int_0^d dz P(z)$ 
 for $P_b=0$  from eq 41.
In our case,   the integral of $P(z)/\mu_0$ 
in the first layer is  $1.6\times (-0.26)= -0.42$ nm$^{-2}$ 
and that in the second layer is $8.2\times 0.073=0.60$ nm$^{-2}$. 
As a result, we obtain  $\Phi_{00}^{\rm w}=-0.09$ V. 
However,  with increasing the water adsorption, 
 the first layer contribution increases, leading to positive  
 $\Phi_{00}^{\rm w}$.  In fact, if we increased the coefficient 
$C_3$ of the attractive part of the wall potential in eq 4 by 10 times 
(with the other parameters unchanged), 
 we obtained $\Phi_{00}^{\rm w}=0.2$ V. Previously,
   Willard {\it et al.} \cite{Madden1} 
found   $\Phi_{00}^{\rm w}\sim 0.8$ V for  strong adsorption.

\vspace{2mm}
{\bf C. Hydrogen bonds}. We also examine 
 the intermolecular hydrogen bonds, which 
have been defined in various manners
\cite{Rao,Luzar1,Kumar,Zie}. We  treat 
 two  molecules  to be hydrogen-bonded 
if one of the intermolecular OH distances 
is shorter  than 2.4 $\rm \AA$ 
and the angle between the OO vector 
and one of their intramolecular  OH bonds is smaller than 
$\pi/6$.  A similar  definition was used by  Zielkiewicz\cite{Zie}.   
We here  consider  the  hydrogen-bond number $m_k$ from O and H atoms 
of  molecule k 
and the hydrogen-bond angle $\theta_{k{\rm OH}}=\cos^{-1}( 
 {Z}_{k{\rm OH}}/| {\bi R}_{k{\rm OH}}|)$ 
with respect to the $z$ axis, where ${\bi R}_{k{\rm OH}}$ 
is the OH  bond vector  
(with $Z_{k{\rm OH}}$ being its $z$ component) 
from O of molecule k to H of another molecule. 
%For the tetrahedral structure,   $m_k$  is ideally 4 for each k  
%(two for its oxygen atom and two for its protons). 

In Fig.5(b),  we display  a snapshot  of the 2D molecular positions 
$(x_{k{\rm G}},y_{k{\rm G}})$   and the associated  intermolecular hydrogen 
bonds  with $1\le m_k\le 5$ 
in the bottom Stern layer for $\Delta\Phi=0$.   
The  average of  $m_k$ is 3.2 here\cite{Rossky}, 
which is smaller than their bulk average 3.6. 
We can see  marked heterogeneous  clustering of the hydrogen 
bonds.  In particular, in blank regions,  the molecules 
 are collectively lifted by about 1 $\rm \AA$ 
due to the hydrogen bonding near the wall.  
These molecular configurations evolve on a timescale of 1 ps\cite{Ber2}. 
In Fig.5(c), we examine  the equilibrium distribution  
of $\cos\theta_{k{\rm OH}}$   at the bottom 
for $\Delta\Phi=0$, which is  maximized at angle 
 $1.1\pi/2$ for  the first layer and 
 $0.73\pi/2$ for  the second. As was reported previously
\cite{Madden1,Rossky,Ber2}, 
 the intermolecular hydrogen bonds 
%near the walls incorrect
near a  wall 
tend to make relatively small angles with respect to 
 the wall plane.
% because of the dipole alignment along the $z$ axis.  
In Fig.5(d), we present the 2D structure factors 
of the molecular positions for $\Delta\Phi=0$ in the three regions 
 given by $z<d$, $|z-H/2|<1.4$  $\rm \AA$, and  $H-z<d$.
For small $k<1$ ${\rm \AA}^{-1}$,   
density fluctuations  are enhanced 
appreciably  near the walls, 
%but are  not detectable  in the middle. incorrect
but enhancement is not detectable in the middle.

% Lee {\it et al.} \cite{Rossky} 
%examined the surface  molecular configurations  and 
%found a considerable 
%decrease in the hydrogen bond number near  a hydrophobic wall.  
  Bratko {\it et al.}\cite{Luzar}  applied 
electric field to water  in the  directions   
   parallel and perpendicular  to   hydrocarbon-like walls. 
They found that   electrostriction  
and   surface affinity (relevant  to electrowetting)  
were more pronounced  for parallel field 
than  for  perpendicular field as a result of 
  hydrogen bond optimization.

\vspace{2mm}
{\bf D. Local Electric Field}. 
In  Appendix A,  we will define  the local electric field 
${\bi E}_k^{\rm loc}$  acting on each molecule $k$  as 
 a linear combination of ${\bi E}_{k\alpha}$  ($\alpha={\rm H1, H2},  {\rm M}$)  at the constituting  charged points. 
  For the TIP4P$/$2005 model, it is expressed as  
\be
{\bi E}_k^{\rm loc}
 =
\frac{1}{2}({1+b_{\rm M}})({\bi E}_{k{\rm H1}}+ {\bi E}_{k{\rm H2}}) 
-b_{\rm M} {\bi E}_{k{\rm M}},
\en 
where  $b_{\rm M}= 0.208$. If we write  ${\bi E}_k^{\rm loc}= 
\sum_\alpha A_\alpha {\bi E}_{k\alpha}$, 
we have 
$A_{\rm H1}=A_{\rm H2}=(1+b_{\rm M})/2$ and $A_{\rm M}= -b_{\rm M}$.  
Using eq 48, we calculated the average  of its  $z$ component 
$E_{\rm loc}= \av{ {E}_{zk}^{\rm loc}}_b$. 
We also calculated  the bulk average field $E_b$ 
and polarization $P_b$ in eq 37. 
They are related as  
\be 
E_{\rm loc}= 
\av{ {E}_{zk}^{\rm loc}}_b = E_b + 4\pi \gamma_{\rm loc}P_b, 
\en 
which is the definition of the Lorentz factor  $ \gamma_{\rm loc}$. 
In our simulation,  
we find  $ \gamma_{\rm loc}\cong 0.58$ for  
all  $\Delta\Phi$ investigated (see  Table 1  
for $\Delta\Phi=0.19$ and 1.9 V). 
Hereafter, $\av{\cdots}_b$ denotes 
 the averages   over  molecules 
in the  region $0.3H<z_{k{\rm G}}<0.7H$ 
and over a time interval of 6 ns, 
where $z_{k{\rm G}}$ is the $z$ component of 
the center of mass  of molecule $k$.

We further  consider   the  equilibrium   distribution 
of  the $z$ component  
 ${  E}_{zk}^{\rm loc}$ in the bulk  defined by   
\be 
P_{\rm loc}(E)=\av{  \delta({E}_{zk}^{\rm loc}-E)}_b, 
\en 
In Fig.6(a), the distribution $P_{\rm loc}(E)$ of the 
local field  along the $z$ axis 
is non-Gaussian   with broad width about $15$ V$/$ nm.
To understand this width, 
we  note  that  a charge $e$ separated by $\sigma(\cong 3.2$ $\rm\AA$)  
creates an electric field with size $e/\sigma^2\cong  14$ V$/$nm. 
%In terms of $ P_{\rm loc}(E)$, 
The mean value     is written as 
$
E_{\rm loc}= 
\av{{E}_{zk}^{\rm loc}}_b =\int dE P_{\rm loc}(E) E.
$  
where  $E_{\rm loc}= 
0.53$  and 4.7 V$/$nm 
for $\Delta\Phi=0.19$ and  1.9 V, respectively.

The  distributions of  electric  field  fluctuations 
  have been  calculated in water   and  electrolytes 
(for $\Delta\Phi=0$)\cite{Gei,Dellago,Ka1}. 
%play  crucial roles  in dissociation reactions 
%and vibrational spectroscopic response.
In particular, Sellner {\it et al.}\cite{Ka1} have obtained 
those at O and H sites, which  resemble  to $P_{\rm loc}^z(E)$ 
 for  $\Delta\Phi=0$ in Fig.6(a).  
We  remark that  the effect of 
molecular stretching in strong local 
 field should be examined \cite{Yu,polarizable}.

\vspace{2mm}
{\bf E. Continuum  Electrostatics}. 
Next, for each  molecule $k$ 
in the region  $0.3H<z_{k{\rm G}}< 0.7H$,  
%Following the classical theory of dielectrics\cite{Fro,Onsager,Kirk}, 
 we  consider  a sphere with radius $R=4\sigma=12.6$ ${\rm \AA}$  
around the center of mass   ${\bi r}_{k{\rm G}}$. We then 
 divide ${\bi E}_{k}^{\rm loc}$ as   
\be 
{\bi E}_{k}^{\rm loc}= {\bi E}_{k}^{\rm in}+ {\bi E}_{k}^{\rm out},
\en  
where the first term is  the   contribution 
from molecules  $\ell$  inside the  sphere 
($|{\bi r}_{\ell{\rm G}}-{\bi r}_{k{\rm G}})|<R$)
and the second term is that from those outside it 
and  the surface (or image) charges. 
Using   the step function $\theta(u)$, 
we write ${\bi E}_k^{\rm in}$  as    
\be 
{\bi E}_k^{\rm in}=
{\sum_{{\ell\neq k},\alpha,\beta}}
 \theta(R-|{\bi r}_{\ell{\rm G}}-{\bi r}_{k{\rm G}})|) A_\alpha 
q_{\beta} {\bi g}({\bi r}_i- {\bi r}_{j}),
\en 
where  $i=k\alpha$,   $j=\ell\beta$  with 
$\ell \neq k$ and $\alpha,\beta=$
H1, H2, M, and  ${\bi g}({\bi r})=
 r^{-3}{\bi r}$. The coefficients $A_\alpha$  appear below eq 48. 

In the  continuum electrostatics\cite{Fro}, 
   ${\bi E}_k^{\rm out}$ 
is given by the classical value  
$ E_b+ 4\pi P_b/3= [(\ve+2)/3]E_b $ 
along the $z$ axis in 
the 1D   geometry.
This approximation  should become  increasingly accurate 
with increasing $R$.     
%Here, setting for $R=4\sigma$, 
Thus, we consider  the deviation of ${\bi E}_k^{\rm out}$ from 
its continuum limit, which is  written as        
\be 
{\bi F}_k^{\rm loc}= 
{\bi E}_k^{\rm out}- (E_b+ 4\pi P_b/3){\bi e}_z. 
\en 
We calculated the equilibrium distribution  
 of its $z$ component  ${ F}_{zk}^{\rm loc}$ in the bulk:      
\be 
P_{\rm L}(E)=\av{ \delta({F}_{zk}^{\rm loc}-E)}_b.
\en 
In Fig.6(b), $P_{\rm L}(E)$  is excellently  Gaussian:     
$
P_{\rm L}(E)\cong \exp(-E^2/2 s_{\rm L})/\sqrt{2\pi s_{\rm L}},
$  
where $\sqrt{s_{\rm L}} = 0.45$, 0.42, and 0.35 V$/$nm 
 for $\Delta\Phi= 0$, 1.9, and 3.8 V, respectively. 
The mean value from  $P_{\rm L}(E)$ vanishes,  
so $F_{\rm loc}= \av{{F}_{zk}^{\rm loc}}_b \cong 0$. 
See Fig.7(b) also, 
where $F_{\rm loc}\cong 0$ for any  $\Delta\Phi$ investigated. 
Since $E_{\rm loc}= \av{{ E}_{zk}^{\rm loc}}= 
\av{ E_{zk}^{\rm in}} + E_b+4\pi P_b/3$ from eq 51,
  the Lorentz factor  $\gamma_{\rm loc}$ in eq 49  is expressed as  
\be 
\gamma_{\rm loc}= 1/3+ E_{\rm in} /4\pi P_b . 
\en 
where $E_{\rm in}= \av{E_{zk}^{\rm in}}_b$. Thus, the deviation of 
$\gamma_{\rm loc}$ from the classical value $1/3$ arises from 
the molecules inside the   sphere. Our calculation 
of $E_{\rm loc}$ gives 
   $E_{\rm in}
\cong 0.25\times 4 \pi P_b$. 
Therefore, the  internal field fluctuations ($\sim e/\sigma^2)$ 
are due to   the contribution  ${\bi E}_k^{\rm in }$ 
from the molecules within the sphere. 
As a result, the distribution of 
 ${\bi E}_k^{\rm in }$ is nearly equal to 
that of  ${\bi E}_k^{\rm loc }- (E_b+4\pi P_b/3){\bi e}_z$. 
 in the bulk. We confirmed that the distribution  of its $z$ component 
$\av{\delta( E_{zk}^{\rm in }-E)}_b$ is equal to $P_L(E+ E_b+4\pi P_b/3)$ 
 (see Fig.6(a)).

The above  results  indicate that  the continuum electrostatics can 
provide an accurate  approximation for   
the electric field  ${\bi E}_k^{\rm out}$ 
from dipoles  and charges outside a 
nanometer-size sphere for  each molecule $k$ in the bulk. 
We should further check  this aspect   with varying $R$ 
and for other geometries.

%6
\begin{figure}
\includegraphics[width=0.96\linewidth]{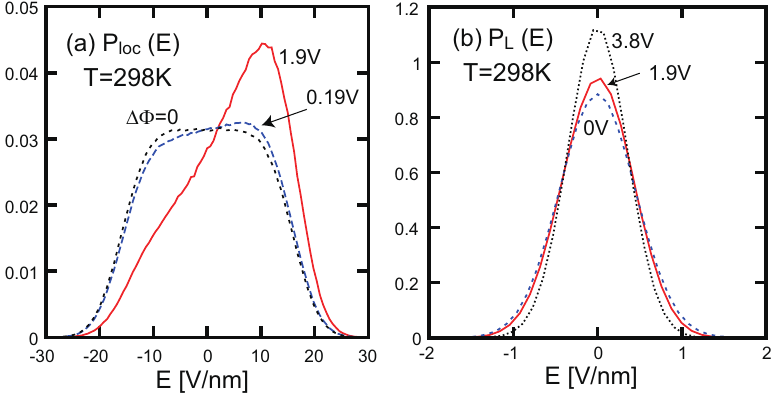}
\caption{   
(a) Distribution 
 $P_{\rm loc}(E)$ of local field ${E}_{zk}^{\rm loc}$   
in  eq 50  for 
$\Delta\Phi=0$, 0.19, and 1.9 V ($E_a=0$, 0.042, 
and 0.42 V$/$nm). 
(b) Distribution   $P_{\rm L}(E)$ of  deviation ${F}_{zk}^{\rm loc}$ 
in eq 54  for $\Delta\Phi=0$, 1.9, and 3.8 V ($E_a=0$, 0.42, 
and 0.84 V$/$nm), which is  Gaussian with narrow width. } 
\end{figure}

%7
\begin{figure}
\includegraphics[width=0.96\linewidth]{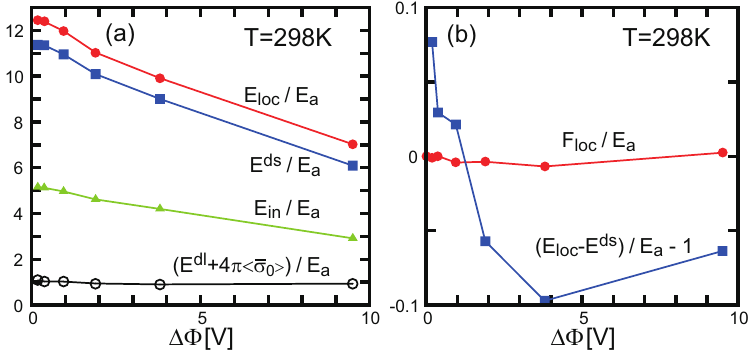}
\caption{   Average fields 
along the $z$ axis divided by $E_a$ vs $\Delta\Phi$  in bulk.  
In (a),  displayed  are   ${ E}_{\rm loc}=\av{{ E}_{zk}^{\rm loc}}_b$, 
${ E}_{\rm in }= \av{E_{zk}^{\rm in}}_b$, 
${ E}^{d\ell}+4\pi\av{{\bar\sigma}_0} = 
\av{{ E}_{zk}^{d\ell}}_b +4\pi\av{{\bar\sigma}_0}$,
and ${ E}^{ds}=\av{{ E}_{zk}^{ds}}_b $. 
In (b), combinations 
$F_{\rm loc}=\av{{ F}_{zk}^{\rm loc}}_b  = ({ E}_{\rm loc}- 
{ E}_{\rm in }) - (E_b+4\pi P_b/3) $   and 
${ E}_{\rm loc}- { E}^{ds}- E_a$  are shown to be  small. 
  }
\end{figure}

\vspace{2mm}
{\bf F. Average Electric Fields vs $\Delta\Phi$}. 
We  also examine   
long-range and short-range parts 
of  the dipolar field ${\bi E}_i^d $  
 on charge $i$ in eq 25   
using the potential division  $1/r=\psi_\ell(r)+\psi_s(r)$ in eq 11. 
Setting ${\bi g}_\ell({\bi r})= -\nabla\psi_\ell(r)$ 
and  ${\bi g}_s ({\bi r})= -\nabla\psi_s(r)$, we express  these two parts  as  
\bea 
&&{\bi E}_i^{d\ell} =  \sum_{ {\bi m}_\perp} 
 {\sum_{j}}  
{ q_j} {\bi g}_\ell({{\bi r}_i- {\bi r}_{j} +  L{\bi m}_\perp}) , 
\\
&&{\bi E}_i^{ds} =   \sum_{ {\bi m}_\perp} 
 {\sum_{j}}' { q_j}{\bi g}_s({{\bi r}_i- {\bi r}_{j} +  L{\bi m}_\perp}) .  
\ena 
Then,  ${\bi E}_i^{d}= 
{\bi E}_i^{d\ell}+ {\bi E}_i^{ds} $. 
Notice that the self terms ($j=i$ for ${\bi m}_\perp ={\bi 0}$)  
can be included  in the long-range part in eq 56, because 
${\bi g}_\ell({\bi 0})={\bi 0}$ from $\psi_\ell(r)
= (2\gamma/\sqrt{\pi})(1- \gamma^2 r^2/3+\cdots)$ 
for small $r$.  Thus, we may rewrite ${\bi E}_i^{d\ell}$ as 
\be 
{\bi E}_i^{d\ell} =  \sum_{ {\bi m}_\perp} \int d{\bi r}' 
\rho ({\bi r}')  {\bi g}_\ell({{\bi r}_i- {\bi r}' +  L{\bi m}_\perp}) , 
\en 
where $\rho({\bi r})=\sum_j q_j\delta({\bi r}-{\bi r}_j)$ 
is the charge density in the cell. 
These two parts  contribute to the  
 local field ${\bi E}_k^{\rm loc}$ in eq 48 
as $ 
 {\bi E}_{k}^{d\ell} = 
\sum_\alpha A_\alpha{\bi E}_{k\alpha}^{d\ell}$ and 
$  {\bi E}_{k}^{ds} 
= \sum_\alpha A_\alpha{\bi E}_{k\alpha}^{ds}.
$

To  estimate  the long-range part ${\bi E}_{i}^{d\ell}$, 
we   consider 
its continuum limit ${\bi E}^{d}_{\rm con}({\bi r}_{i})$. 
  The latter  is obtained if  we replace  
$\rho({\bi r})$ by $ P_b[\delta(z-H+d)-\delta (z-d)]$ 
in eq 58, since the polarization 
is given by $P_b{\bi e}_z$ 
outside  the Stern layers with  thickness $d$ 
in the continuum description.  Then, 
the space integral  
 $\int d{\bi r}'$ is performed to give 
\be 
{\bi E}^{d}_{\rm con}({\bi r})= -4\pi P_b {\bi e}_z 
=(E_b-4\pi\av{{\bar\sigma}_0}){\bi e}_z,
\en
where $z\gg d$ and $H-z\gg d$. 
We  numerically checked  that 
${\bi E}_i^{d\ell}$ is  in fact   close to  ${\bi E}^{d}_{\rm con}$ in eq 59 
for $i$ in the bulk. This means that the  molecules 
near the walls give rise to a dominant contribution to 
${\bi E}_i^{d\ell}$.   It follows that   the sum  of the averages 
$\av{{ E}_{zk}^{d\ell}}_b + 4\pi \av{{\bar\sigma}_0}$ 
is   of order $E_b\sim E_a$. 
In eq 24, we notice  that the last  term 
$4\pi{{\bar\sigma}_0}{\bi e}_z$ 
is largely canceled by the long-range part of 
the first term ${\bi E}_i^{d}$.

In Fig.7, we  show  the bulk averages of the $z$ components 
of  ${\bi E}_k^{\rm loc}$, ${\bi E}_k^{\rm in }$, 
${\bi E}_k^{d\ell}$, and ${\bi E}_k^{ds}$  (divided by $E_a$) 
%for the molecules in the bulk ($0.3H<z_{k{\rm G}}<0.7H)$ 
as functions of  $\Delta\Phi$.
% which are  written   as 
%${ E}_{\rm loc}$, ${ E}_{\rm in }$, ${ E}^{d\ell}$,
%and ${ E}^{ds}$, respectively. 
In (a), $E_{\rm loc}=\av{{ E}_{zk}^{\rm loc}}_b$ 
 and $E^{ds}=\av{{ E}_{zk}^{d s}}_b$ are of order $10E_a$. 
In accord with the discussion below eq 59, 
we can see the following relations, 
\bea 
&&{ E}^{d\ell}=\av{{ E}_{zk}^{d \ell}}_b 
 \cong -4\pi  \av{{\bar\sigma}_0}+E_a, \\
&&{ E}^{ds}=\av{{ E}_{zk}^{d s}}_b 
\cong { E}_{\rm loc}- E_a,
\ena 
which are consistent with eq 24  if ${\bi E}_s$ is neglected. 
Furthermore, in (b),  $F_{\rm loc}=\av{{ F}_{zk}^{\rm loc}}_b=
( { E}_{\rm loc}- { E}_{\rm in })- (E_b+4\pi P_b/3) $ 
is almost zero  for any $\Delta\Phi$ 
in accord with Fig.6(b), 
while ${ E}_{\rm loc}- { E}^{ds}- E_a= E^{d\ell}+4\pi 
 \av{{\bar\sigma}_0}-E_a $ 
is at most $10\%$ of $E_a$ supporting the discussion below eq 59.

\section{Field Reversal}

{\bf A. Situation and Method}.
In this section, we  examine   the relaxation after 
field reversal at $T=298$ K  in the $NVT$ ensemble 
to avoid heating. 
For simplicity,   the applied potential 
  difference $\Delta\Phi$ (field $E_a$) is changed  instantaneously 
from $-1.9$ V ($-0.42$ V$/$nm) to 1.9 V  ($0.42$ V$/$nm) at $t=0$. 
The  system was in equilibrium for $t<0$ 
and transient  behaviors  follow for $t>0$.  
The polarization $M_z(t)$ changes continuously  from $M_z(0)<0$ 
to $M_z(\infty)=- M_z(0)$.   From the last term in  eq 23,  
the electrostatic  energy $U_{\rm m}$ decreases by  $2E_a |M_z(0)|$ 
after the field reversal.  In this paper, 
we thus present the results in  the $NVT$ ensemble 
to suppress heating. If we used the $NVE$ ensemble, 
we found heating by $\Delta T= 10-15$ K, where 
the kinetic energy increase given by   $3k_B\Delta T$ per molecule 
was about one third of   $2E_a |M_z(0)|/N$. 
However, these two simulations 
yielded  essentially the same  microscopic  reorientation dynamics.

As an example,  in Fig.8, we write   
 one molecule at the center of each  panel,  
%at  $t=0.45$, 0.54, 0.63, and 0.72 ps,  
which  is undergoing  a large angle change.  
The other molecules depicted are those which  have 
been connected to the center molecule by hydrogen-bonding 
at some $t$  in the  time range  0.45 ps $\le t\le$ 0.72 ps. 
These surrounding molecules 
are  also rotating in complex manners. 
In this section, we use the definition 
of hydrogen bonds given in Sec.IIIC. 
In  the literature  \cite{Ohmine,Oh1,Laage,La1}, 
  large-angle changes have been reported to occur 
cooperatively  as  reorganization of 
 the hydrogen bond network. 
%7
\begin{figure}
\includegraphics[width=0.96\linewidth]{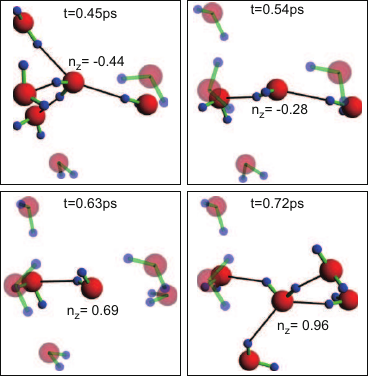}
\caption{ 
An example of reorientation dynamics of 
 water molecules  at $t=0.45$, 0.54, 0.63, and 0.72 ps 
in one run of  field reversal,  where the $z$ 
axis is in the vertical direction. 
Intermolecular hydrogen bonds are also written (black bars). 
 To a molecule at the center, value of $n_z=\cos\theta$ 
is  attached in  each panel.  The other molecules have 
been connected by hydrogen-bonding at some $t$ 
in time interval $[$0.45 ps,0.72 ps$]$.
 }
\end{figure}

In  water at room temperatures, 
the thermal fluctuations are large. 
Thus, we performed 50 independent runs  of 60 ps length 
to produce Figs.9, 10,  and 13  and the bottom panels of Fig.12. 
In this section, 
 $\av{\cdots}$ denotes taking  the nonequilibrium 
average over these 50 runs, which should not be confused 
with the time average in the previous section.

{\bf B. Results of 1D Profiles}.  
In  Fig.9, we show  1D profiles of $P(z,t)$ in 
eq 30  and $\Psi(z,t)$ in eq 31  at $t=0.001$, 1.5, and 6 ps.
Here, the system  is divided into two Stern layers and a bulk 
region   even in transient states. In the bulk region, $P_b(x,t)$ increases 
continuously, while   $E_b(t)$ increases from 
a negative value ($t<0)$ to a positive value 
discontinuously at $t=0$ and then decreases 
to the final  positive value. 

In Fig.10, , we give  $\av{{\bar{\sigma}}_0}(t)$,  
$ P_b(t)$,  and  $\av{M_z}(t)/V$ vs $t$ in (a). 
Here, eqs 16  and 34  hold, so $4\pi \av{{\bar\sigma}_0}(t)$  increases 
  by $2\times 0.42$ V$/$nm discontinuously  at $t=0$.
These quantities are very close since $E_a$ and 
$E_b(t)$ are much smaller.  In  (b), we give the 
average of the $z$ component of the polarization 
direction   $ {\bar n}_{k}(t)$ in eq 1, denoted by $\av{n_z}(t)$,  
for the molecules in the three regions 
$z_{k{\rm G}}<0.1H$, $z_{k{\rm G}}>0.9H$, and $0.1H<z_{k{\rm G}}<0.9H$ 
separately. The relaxations  near the walls and in the bulk 
are similar because the   adsorption is weak 
in our simulation. Also displayed are $E_b(t)$ in (c) and 
 $\Phi_{0}^{\rm w}(t)$ and   $\Phi_{H}^{\rm w}(t)$ in (d), 
which  become small with 
considerable fluctuations  for $t>5$ ps.  
The quantities in (a)-(d) relax exponentially with a common relaxation rate 
$\tau_r^{-1}$ with  $\tau_r=2.84$ ps at long times.  
In particular, in the inset in (a), this exponential decay can be 
seen from the beginning  in the average polarization  difference    
$
\Delta \av{M_z} (t) =\av{M_z}(\infty) - 
\av{M_z}(t). 
$  

%9
\begin{figure}
\includegraphics[width=0.96\linewidth]{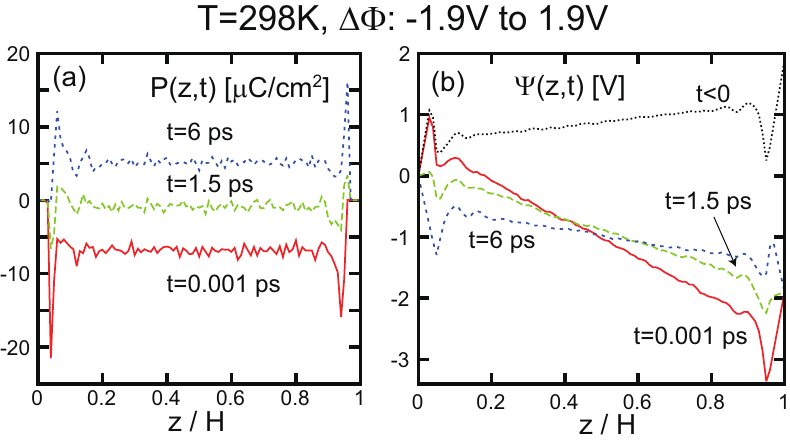}
\caption{ Space-time-evolution after  field reversal 
 from $\Delta\Phi=-1.9$ V to 1.9 V at $T=298$ K.  (a) $P(z,t)$ 
and (b) $\Psi(z,t)$ at  $t=0.001$, 1.5, and 6 ps.
These are  averages over 50 runs. }
\end{figure}
%10
\begin{figure}
\includegraphics[width=0.96\linewidth]{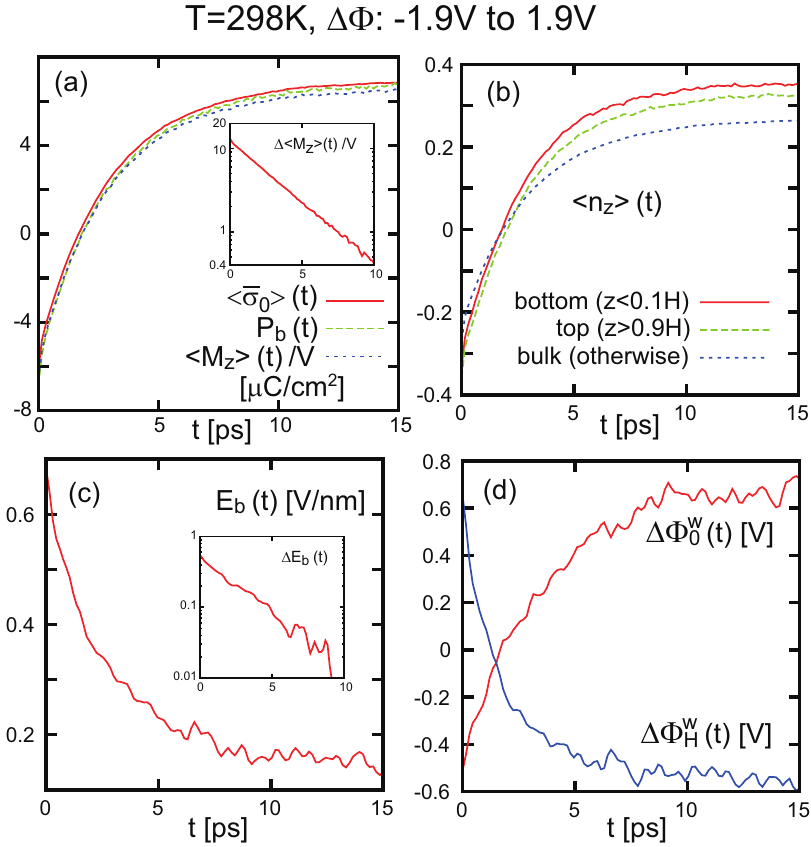}
\caption{ Time-evolution after  field reversal 
 from $\Delta\Phi=-1.9$ V to 1.9 V at $T=298$ K.   (a) 
$\av{{\bar{\sigma}}_0}(t)$,  $ P_b(t)$,  and  $ \av{M_z}(t)/V$ vs $t$. 
Inset:  $\Delta \av{M_z}(t)=  \av{M_z}(\infty)- \av{M_z}(t)$ 
divided by $V$ 
on a semi-logarithmic scale. 
 (b) $\av{n_z}(t)$ 
(average of  $ {n}_{zk}(t)$) 
vs $t$  for molecules in  three regions:  
$z_{k{\rm G}}<0.1H=4.5$ ${\rm \AA}$, 
$z_{k{\rm G}}>0.9H$, and $0.1H<z_{k{\rm G}}<0.9H$. (c) $E_b(t)$ vs $t$. 
Inset: $\Delta E_b(t)=E_b(t)-E_b(\infty)$ on a semi-logarithmic scale. 
(d) $\Delta\Phi_{0}^{\rm w}(t)$ and   $\Delta\Phi_{H}^{\rm w}(t)$ in eq 40  vs $t$.  These are  averages over 50 runs.
 }
\end{figure}
%11
\begin{figure}
\includegraphics[width=0.96\linewidth]{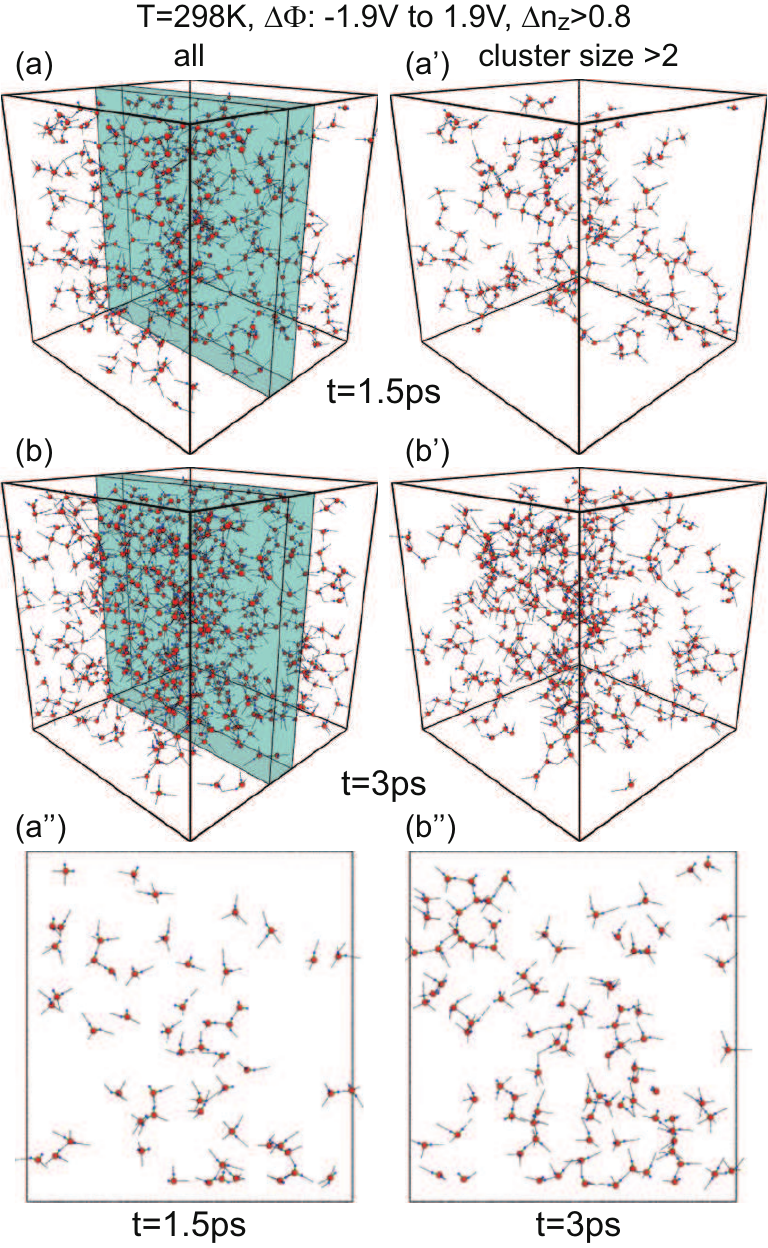}
\caption{ 
Largely rotated (LR) molecules  with  big rotational jumps 
satisfying eq 62  at (a) $t=1.5$ and (b) 3 ps (left), 
 among which selected  are those  forming hydrogen-bonded clusters 
with sizes  exceeding  2 at (a') $t=1.5$ and (b') 3 ps (right). 
The $z$ axis is in the vertical direction. 
Cross-sectional snapshots at  (a'') $t=1.5$ and (b'') 3 ps (bottom), 
where displayed are molecules with their centers of mass 
in the shaded region  with thickness $6~{\rm \AA}$ in (a) 
and (b). 
}
\end{figure}

%12
\begin{figure}
\includegraphics[width=0.96\linewidth]{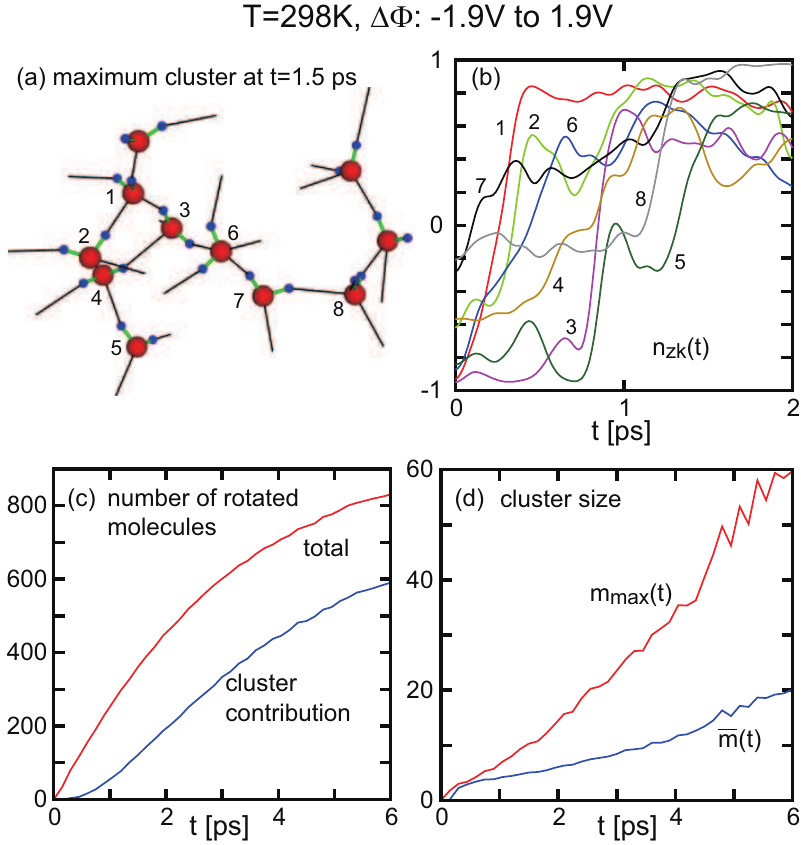}
\caption{   
(a) Snapshot of a hydrogen-bonded 
cluster whose members undergo big rotational jumps. 
(b) Time-smoothed trajectories   ${\bar n}_{zk}(t)$ in eq 63
for  8  molecules in (a). 
(c) Number of largely rotated molecules  (total) and that 
of  largely rotated molecules belonging to H-bonded 
clusters with sizes exceeding  2 (cluster contribution) vs $t$. 
(d) Average  maximum cluster size $m_{\rm max}(t)$ 
and  mean  cluster size ${\bar m}(t)$ vs $t$. 
Curves in (c) and (d) are obtained as  averages over 50 runs. 
 }
\end{figure}

{\bf C. Big Rotational Jumps and Hydrogen-Bonded Clusters}. 
In water, large  orientational  changes of the dipoles 
are accompanied by  breakage and reorganization 
of the hydrogen-bond network even at room temperatures \cite{Laage,La1}.  
This suggests that large angle changes 
are relevant in the relaxation 
 after a  reversal of strong applied field.
%can in fact  be seen in Fig.8 and  Fig.12(b) below. 
Thus, we consider   largely rotated (LR) molecules with 
 angle jumps  determined by 
\be 
n_{zk}(t)-n_{zk}(0)>0.8,
\en  
where $n_{zk}(t)=\cos[\theta_k(t)]$ is  the $z$ component 
of the polarization direction ${\bi n}_k(t)$ in eq 1  
with $\theta_k(t)$ being the angle of the polarization 
with respect to  the $z$ axis.

In Fig.11, we present snapshots 
of the LR molecules satisfying 
eq 62 in the left panels at $t=1.5$ 
and 3 ps. Among these molecules, we pick up  
those belonging to hydrogen-bonded clusters 
with member numbers exceeding  2 in the right panels. 
In the bottom panels, cross-sectional snapshots 
at  these times are  displayed. 
In these snapshots,  large-angle  rotational jumps   occur collectively 
and heterogeneously. We show that 
the LR molecules themselves 
 form hydrogen-bonded clusters, which we have   detected   
by  comparing two sets of  the orientation configurations at two times. 
Such  dynamic heterogeneities  
have been detected in translation and rotation 
in supercooled water \cite{Stanley,Mazza}  
 and in double glass\cite{double}.

In Fig.12, we illustrate  a cluster composed of 11
  LR molecules in (a), which is the largest one 
at $t=1.5$ ps in one simulation run.  Here,   
the  orientations  ${\bi n}_{k}(t)$ 
of these molecules exhibit  rapidly varying  thermal 
fluctuations, so we consider   the  
time average of their $z$ components,  
\be 
\bar{n}_{zk}(t)= \int_{t-\Delta t}^t dt' n_{zk} (t')/\Delta t. 
\en   
In (b), setting  $\Delta t= 0.015$ ps, 
we plot $\bar{n}_{zk}(t)$ vs $t$ for 8  members  depicted in (a).
  We can see these molecules 
undergo simultaneous 
big rotational jumps  in a time interval of 1 ps. 
In (c), as functions of $t$, 
we show the total number 
of the LR molecules and that 
of the LR molecules forming  hydrogen-bonded 
clusters with sizes  exceeding  2.  
In (d), we plot the average  maximum cluster size 
$m_{\rm max}(t) $ 
and  the mean  cluster size,
\be 
{\bar m}(t)= 
\av{\sum_{m>2} N_m(t) m^2 /\sum_{m>2} N_m(t) m}, 
\en 
where $N_m(t)$ is  the number of the clusters 
with size $m$,

{\bf D. Big Jump Fraction and Hydrogen-Bond Numbers}.
We  need  to  quantitatively  show  how 
big-jump reorientations can be a dominant mechanism of the relaxation. 
To this end, we define   the    big-jump fraction by  
\be 
\phi_M^{\rm jump}(t)= 
 \AV{\sum_{k \in {\rm jump}} [n_{zk}(t)-n_{zk}(0)]}
\frac{\mu_0}{ \Delta M(t)} ,
\en 
where we  sum over molecules   $k$ 
satisfying eq 62. In the denominator, we set 
$\Delta {M} (t)  = \mu_0
 \av{\sum_{k} [n_{zk}(t)-n_{zk}(0)]}$
 summing   over all $k$, 
so it  is equal to $\av{M_z}(t)-\av{M_z}(0)$ 
 (see Fig.11(a) for its time dependence). 
In  Fig.13(a), $\phi_M^{\rm jump}(t)$  is small 
at very short times but soon exceeds 0.6 for $t>1 $ ps. 
Therefore,    the relaxation 
is governed by   big rotational jumps  for $t>1$ ps. 

The  reorganization of the hydrogen bond network  
occurs   very rapidly, so   it is a difficult 
task to capture the dynamics 
quantitatively\cite{La1,Luzar1}. 
 Here, we consider  the hydrogen bond number $m_k$ 
for  molecule $k$, which  is ideally 4 (two for its oxygen 
atom and two for its protons)  for the tetrahedral structure.  
Due to structural disorder, however,  its thermal average 
becomes  $\av{m_k}\cong 3.6$ at $T=298$ K  
for our definition of hydrogen bonds. 
In Fig.13(b), we  plot the fractions 
$\phi_{\rm HB}^{\rm jump}(m, t)$ 
 of  the LR molecules with $m_k= m$ 
at time  $t$. They are expressed as 
\be 
\phi_{\rm HB}^{\rm jump}(m, t)= 
 \AV{\sum_{k \in {\rm jump}}\delta_{m m_k(t)}/N_{\rm jump}(t)},
\en 
where $N_{\rm jump}(t)$ is the number of the LR molecules 
satisfying eq 62  at time $t$.  We can see that 
$\phi_{\rm HB}^{\rm jump}(3, t)$ and 
$\phi_{\rm HB}^{\rm jump}(2, t)$ 
decay from 0.40 and 0.20 
at $t=0.4$ ps 
to  the equilibrium values $\phi_{\rm HB}(3)=0.31$ 
and  $\phi_{\rm HB}(2)=0.09$, respectively,   
with a relaxation time about  2 ps. 
There is almost no change  in $\phi_{\rm HB}^{\rm jump}(5, t)$ 
from the equilibrium value $\phi_{\rm HB}(5)=0.060$. 
Note that $\phi_{\rm HB}(4)\cong 0.54$  in equilibrium.
Thus, the nonequilibrium 
distribution $\phi_{\rm HB}^{\rm jump}(m, t)$ 
rapidly approaches   the equilibrium distribution 
$\phi_{\rm HB}(m)$ on a timescale of 2 ps   and 
the deviation $\phi_{\rm HB}^{\rm jump}(m, t)-\phi_{\rm HB}(m)
$ is not large except for very small $t(<0.1$ ps), 
though   a few hydrogen bonds 
are broken  for each big rotational jump.

%13
\begin{figure}
\includegraphics[width=0.96\linewidth]{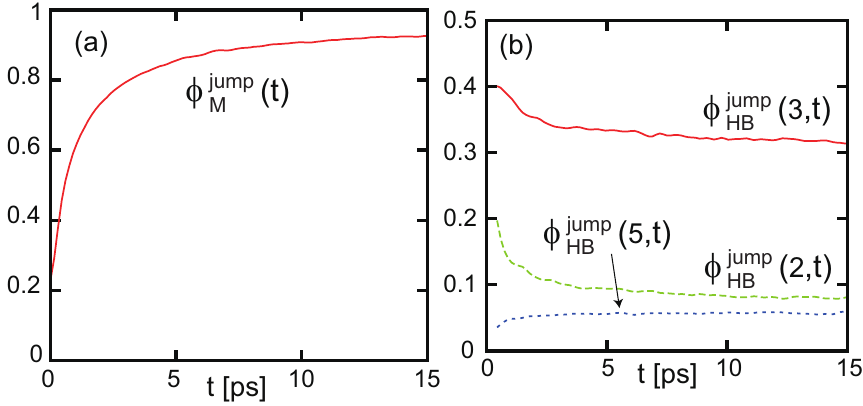}
\caption{  
(a) Big-jump fraction $\phi_M^{\rm jump}(t) $ in eq 65  vs $t$, 
which is the fraction of the contribution 
of LR molecules to $\av{M_z}(t)-\av{M_z}(0)$.
(b) Fractions $\phi_{\rm HB}^{\rm jump}(m,t) $ vs $t$ 
with hydrogen bond number $m=2, 3$ and 5.   
These are  averages over 50 runs. }
\end{figure}

\section{Summary and Remarks}

In this paper, we have  
studied dielectric responses in applied electric field and 
polarization relaxation after field reversal in 
a system of $2400$ water molecules 
 between metal walls at $z=0$ and $H=44.7$$\rm \AA$. 
We have used  the TIP4P$/$2005 model 
and   the 3D Ewald method,  
including  the image effect 
 to realize  the constant potential condition on the walls. 
%Since   real  charges  are those on the  metal surfaces and those  
%in the water molecules, we have focused on the surface charges.  
In the following, 
we summarize our main results with critical remarks.

(i) In Sec.II, we have 
explained our simulation method. We have shown that the surface charges 
 yield  an  electric potential 
consisting of the average part $-4\pi{\bar \sigma}_0 z$ 
and a deviation $\phi_s$, where   ${\bar \sigma}_0$ 
is the mean  surface charge at $z=0$. Then, 
each charge $i$  is acted by  the field from the surface charges 
and  the dipolar field ${\bi E}_i^d$ 
from the other charges in the cell. We have 
 expressed  the electrostatic potential $U_{\rm m}$ 
in terms of $\phi_s$ in eq 23 not using the image charges.  
If $\phi_s$ is negligible, we can justify   the simulation  methods by 
Yeh and  Berkowitz\cite{Yeh} 
and  by Petersen {\it et al.}\cite{Voth}. Since $E_a$ appears linearly in 
$U_{\rm m}$, we can derive the linear response expressions 
such as that for $\ve_{\rm eff}$ in eq 36.  

(ii) In Sec.IIIA,  we have calculated the 2D  
structure factors $S_0(k)$ and $S_H(k)$ 
and the corresponding 2D pair correlation functions  $g_0(\rho)$ and 
$g_H(\rho)$ for the surface charge 
fluctuations at the top and the bottom in Fig.2.  
We have shown that    $\phi_s({\bi r})\to 0$ 
far from the walls  in Fig.3. 
In Appendix C, we have examined how  the 
fluctuation amplitude $e_s(z)= [\av{|\nabla\phi_s|^2}]^{1/2}$ 
decays far from  the walls.  
Therefore,    the bulk properties of water in applied electric field 
are determined by the mean   surface charge 
${\bar \sigma}_0$ \cite{Yeh,Voth}.  
On the other hand, the  molecules near a metal wall are under influence 
of strongly  heterogeneous surface charges. 
Thus, the  molecules near and far from the walls behave  
very differently.

(iii) In Sec.IIIB, we have examined  average 1D profiles     
using the microscopic expression for the polarization 
density ${\bi p}({\bi r})$ in Appendix A, where 
  Stern  layers with thickness 
$d=4.7$ ${\rm \AA}$ and a homogeneous bulk region appear. 
 The ratio  between the  polarization $P_b$ 
and the  electric field $E_b$ in the bulk 
yields  the  dielectric constant $\ve=1+ 4\pi P_b/E_b$. 
The applied field $E_a$  is larger than $E_b$ by  
 $E_a/ E_b= 1+ \ell_{\rm w}/H$, where $\ell_{\rm w}$ 
is a surface electric length about  10 nm. 
Thus,   the dielectric response strongly depends on the cell length $H$.  
In  the previous simulations \cite{Yeh,Yeh1,Hautman,Madden1}, 
 $H$  has been  shorter than $\ell_{\rm w}$.  
%Our  calculated  surface capacitances $C_+$ and $C_-$  are 
%of the same order as  those in  the previous 
%simulations \cite{Hautman,Madden1} but are smaller than the experimental 
%values. 

(iv) 
Furthermore, in Sec.IIIB, we have examined 
 the zero-field surface potential drop $\Phi_{00}^{\rm w}$ dividing    
the Stern layer into two layers. We have 
found that  the polarization is downward (upward)
in the first (second) layer for $\Delta\Phi=0$.
The  H-down orientation in Fig.1 is preferred  in the first layer 
due to the image interaction, 
while a surface-to-bulk crossover takes place  in the hydrogen bonding 
in the second layer. In our case,  the water adsorption is weak and 
 the polarization in the first layer 
is relatively small compared to  that in  the second, 
leading   to   $\Phi_{00}^{\rm w} \sim -0.09$ V. 
 For strong adsorption, a 
positive $\Phi_{00}^{\rm w}$ follows \cite{Madden1}.

(v) In Sec.IIIC, we have  visualized clustering of 
the hydrogen bonds in the Stern layer. 
The hydrogen bond orientations in the Stern layer  
tend to be parallel to the walls as  in Fig.5(c).
Large-scale  density fluctuations  near the  walls 
have also been detected in Fig.5(d). 
These mesoscopic heterogeneities   result  from competition 
between hydrogen bond formation 
and packing near a wall\cite{Rossky,Luzar,Ber2}.

(vi) In Sec.IIID, we have calculated the local field ${\bi E}_k^{\rm loc}$ 
on each molecule $k$ using its microscopic expression  in Appendix B. 
Writing  the  bulk average of its $z$ component 
as $E_{\rm loc}=E_b+ 4\pi \gamma_{\rm loc}P_b$, 
we  have obtained  the Lorentz factor $\gamma_{\rm loc}\cong 0.58$. 
Its deviation  from the classical value $1/3$ is caused by  the 
surrounding nearby molecules.
The local field  exhibits large fluctuations 
with amplitude of order $e/\sigma^2$$\sim 15$ V$/$nm 
 with $\sigma=3.2$ ${\rm \AA}$  
  and its distribution is deformed 
by applied field  as in Fig.6(a).  
We note that the  dipolar energy $\mu_0 e/\sigma^2
=26k_BT$ per  molecule is very large 
and the dipole moment ${\bi \mu}_k$ should 
be mostly along the local field ${\bi E}_k^{\rm loc}$.  
This aspect will  be investigated in future.

(vii) In Sec.IIIE,  we have confirmed that 
the local field contribution from the exterior of a 
nanometer sphere  ${\bi E}_k^{\rm out}$ 
is  given by its  continuum limit $(E_b+4\pi P_b/3){\bi e}_z$  
with small deviations obeying a Gaussian distribution. 
On the other hand, 
the  contribution from the sphere interior ${\bi E}_k^{\rm in}
={\bi E}_k^{\rm loc}- {\bi E}_k^{\rm out}$ exhibits 
 large  fluctuations ($\sim e/\sigma^2)$, where 
relevant is the short-range orientational correlation. 
 In Sec.IIIF, we have  furthermore  divided the dipolar  field 
from the molecules in the cell 
 into long-range  and short-range parts 
as  ${\bi E}_k^{\rm d}={\bi E}_k^{\rm d\ell}+{\bi E}_k^{\rm ds} $
using the Coulomb potential division 
$1/r= \psi_\ell(r)+\psi_s(r)$ in the Ewald method. 
We have found that a main contribution 
to the long-range part ${\bi E}_k^{\rm d\ell}$ 
is produced by the dipoles 
near the walls and   can well be approximated 
by its  continuum limit. 
%give rise to a dominant contribution to .  
%${\bi E}_k^d= {\bi E}_k^{d\ell}+{\bi E}_k^{ds}$.
%${\bi E}_i^{d\ell}\cong -4\pi p_B{\bi e}_z$. 
%The averages of the $z$ components  of the long-range 
%and short-range parts have been calculated to be 
% $ -4\pi {{\bar\sigma}_0}+E_a$ and $ { E}_{\rm loc}- E_a$, respectively, 
%in Fig.7. 
With the   aid of these two   field divisions, we can 
 investigate   how  the continuum description 
can be used in the calculation of the long-range interaction.
%${\bi E}_k^{\rm loc}
%={\bi E}_k^{\rm in}+ {\bi E}_k^{\rm out}$.  
%, where 
%$4\pi {{\bar\sigma}_0}\sim { E}_{\rm loc}\gg  E_a$. 

(viii) In Sec.IV, we have studied   
the orientation dynamics after  field reversal. 
 Due to the presence of the hydrogen bond 
network, the relaxation 
is governed by large-angle  rotational jumps\cite{Laage,La1}, 
as demonstrated in Figs.11-13.  
%where the jump  criterion is   given by eq 62. 
These big  jumps occur in the form of  hydrogen-bonded 
clusters and the resultant dynamic heterogeneity 
 has been  displayed  in Fig.11. 
We have examined how these big jumps 
contribute to the polarization relaxation in Fig.13(a). The threefold 
and fivefold hydrogen bonds 
transiently increase right after big jumps 
as in Fig.13(b). 
Note that  cooperative motions  
   are   more  conspicuous  with smaller thermal 
noises in supercooled water    
 \cite{Stanley,Mazza}. 
Generally in supercooled  anisotropic  liquids, 
 dynamic heterogeneity emerges  both  in rotational  and translational motions    \cite{double}.

\begin{acknowledgments}
This work was supported by KAKENHI 
 (Nos. 25610122  and 25000002).  
The numerical calculations were carried out on SR16000
at YITP in Kyoto University.
\end{acknowledgments}

\vspace{2mm}
\noindent{\bf Appendix A: 
Local Electric Field  in  Water }\\
\setcounter{equation}{0}
\renewcommand{\theequation}{A\arabic{equation}}
%\mu_0=2.305D (��������������3.62)

In the TIP4P$/$2005 model\cite{Vega},  
a water molecule  is treated as a rigid 
isosceles triangle. For  a  molecule $k$, 
we  write the positions 
of its oxygen atom  and two protons 
 as ${\bi r}_{k{\rm O}}$, ${\bi r}_{k{\rm H1}}$,
 and  ${\bi r}_{k{\rm H2}}$, respectively. 
Here, $a_{\rm OH}=|{\bi r}_{k{\rm H1}}-{\bi r}_{k{\rm O}}|=
|{\bi r}_{k{\rm H2}}-{\bi r}_{k{\rm O}}|=0.957~{\rm \AA}$ 
and the angle between ${\bi r}_{k{\rm H1}}-{\bi r}_{k{\rm O}} $ and 
 ${\bi r}_{k{\rm H2}}-{\bi r}_{k{\rm O}}$ is 
$\theta_{\rm HOH}=104.5^{\circ}$, so 
$a_{\rm HH}= |{\bi r}_{k{\rm H1}}-{\bi r}_{k{\rm H2}}|=
2a_{\rm OH}
\sin(\theta_{\rm HOH}/2)=0.844~{\rm \AA}.
$   
For each molecule, the  charges are  at the proton 
positions  with $q_{\rm H} 
%=13.26(\epsilon\sigma)^{1/2} 
= 0.5564e$ and at another position M, 
\be 
 {\bi r}_{k{\rm M}}={\bi r}_{k{\rm O}}+ a_{\rm OM}{\bi n}_{k}
\en 
 with $q_{\rm M}=-2q_{\rm H}$  and   $a_{\rm OM}=0.1546{\rm \AA}$. 
The elementary charge is $ e=23.82 (\epsilon\sigma)^{1/2}$. 
The   ${\bi n}_k$ is   the unit vector 
from ${\bi r}_{k{\rm O}}$  to the midpoint of  the proton positions,  
\be 
{\bar{\bi r}}_{k{\rm H}}= \frac{1}{2} 
({\bi r}_{k{\rm H1}}+{\bi r}_{k{\rm H2}}).
\en 
Thus,   
$
{\bar{\bi r}}_{k{\rm H}}-{\bi r}_{k{\rm O}}=a_{\rm d}
{\bi n}_{k}$ 
with   $a_{\rm d}=a_{\rm OH}\cos(\theta_{\rm HOH}/2)
=0.5859{\rm \AA}$. 
The dipole  is expressed as in eq 1 with  
$ 
\mu_0=2q_{\rm H}(a_{\rm d}-a_{\rm OM})= 
3.62 (\epsilon\sigma^3)^{1/2}= 2.305~{\rm D}.
$   

Next, we  shift the  charge positions 
${\bi r}_j$ infinitesimally by  $d{\bi r}_j$ 
with the molecular shape held unchanged.  
%$d{\bi r}_{\rm H2}$, and $d{\bi r}_{\rm M}$.  
Their    images outside the cell 
are also shifted by the same amounts. 
The change in  the electrostatic energy 
in eq 8   is rewritten at fixed $E_a$ as 
\be 
dU_{\rm m}= 
-\sum_k [{\bi F}_k^e \cdot d{\bi r}_{k{\rm G}}
+{\bi F}_k^r\cdot d{\bi \xi}_k 
+{\bi E}_k^{\rm loc}\cdot d{\bi \mu}_k  ].  
\en 
% in terms of the changes in 
%${\bi r}_{\rm G}$, $\bi \mu$, and 
where we introduce  the center of  mass 
and  the relative positional vector between 
two protons   by 
\bea 
&&\hspace{-1cm}
 {\bi r}_{k{\rm G}}= 
\frac{8}{9} {\bi r}_{k{\rm O}} 
+ \frac{1}{18} ({\bi r}_{k{\rm H1}} 
+{\bi r}_{k{\rm H2}} ),\\
&&  \hspace{-1cm}
{\bi\xi}_k ={\bi r}_{k{\rm H1}}-{\bi r}_{k{\rm H2}}. 
\ena  
For each water molecule $k$, 
the conjugate electric forces to 
${\bi r}_{k{\rm G}}$ and  ${\bi \xi}_k$ are given by 
\bea 
&&\hspace{-10mm}
{\bi F}_k^e = q_{\rm H}( {\bi E}_{k{\rm H1}}+ {\bi E}_{k{\rm H2}}
-2{\bi E}_{k{\rm M}}),\\
&&\hspace{-10mm}
{\bi F}_k^r = \frac{1}{2} {q_{\rm H}}({\bi E}_{k{\rm H1}}-{\bi E}_{k{\rm H2}}).
%&&\hspace{-10mm}
%{\bi E}_k^{\rm loc}=
%\frac{1}{2}(1+b_{\rm M})({\bi E}_{k{\rm H1}}+ {\bi E}_{k{\rm H2}}) 
%-b_{\rm M} {\bi E}_{k{\rm M}},
\ena 
The   ${\bi E}_k^{\rm loc}$ in eq A3 is 
the local electric field on water molecule   $k$, 
which is  conjugate to ${\bi \mu}_k$. 
It is expressed as in eq 48  with $b_{\rm M}= 
({a_{\rm OM}-a_{\rm d}/9})/{(a_{\rm d}-a_{\rm OM})}=0.208$.

\vspace{2mm}
\noindent{\bf Appendix B: 
Microscopic Expressions for Polarization Density and 
Poisson Electric Potential }\\
\setcounter{equation}{0}
\renewcommand{\theequation}{B\arabic{equation}}

We  give a microscopic expression for the 
polarization density  ${\bi p}({\bi r})$   in terms of 
the  charged positions  ${\bi r}_j$ for  
polar molecules (with $\sum_j q_j=0$).  
We assume that   ${\bi p}$   is   
related to  the microscopic charge density  
$\rho({\bi r})$ by 
\be 
-\nabla\cdot{\bi p} = \rho = \sum_j q_j \delta ({\bi r}-{\bi r}_j). 
\en 
We  introduce 
 the following 3D   symmetrized $\delta$-function, 
\be
\hat{\delta} ({\bi r} ; {\bi r}_{1},{\bi r}_2) =
\int_0^1 d\lambda~ \delta ( {\bi r}-\lambda {\bi r}_1  
-(1-\lambda){\bi r}_2),
\en 
where ${\bi r}_1$ and ${\bi r}_2$ are particle positions. 
This $\delta$-function  is nonvanishing only on the line segment connecting 
${\bi r}_1$ and ${\bi r}_2$. It is known to appear in the 
microscopic expression for the local stress tensor\cite{Onukibook}. 
For  the TIP4P$/$2005 model\cite{Vega}, 
using the relation  
$({\bi r}_1-{\bi r}_2)\cdot\nabla 
\hat{\delta} ({\bi r} ; {\bi r}_{1},{\bi r}_2)
= \delta ( {\bi r}-{\bi r}_2) - \delta ( {\bi r}-{\bi r}_1)   
$, we find  ${\bi p}$ explicitly as 
\bea 
&&\hspace{-7mm}
{\bi p}({\bi r})= \sum_k  \frac{q_H}{2} \bigg 
[\hat{\delta}({\bi r} ; {\bi r}_{k{H1}},{\bar{\bi r}}_{k{H}})
-\hat{\delta}({\bi r} ; {\bi r}_{k{H2}},{\bar{\bi r}}_{k{H}})\bigg]
 {\bi \xi}_k \nonumber\\
&&\hspace{1.5cm}
+\sum_k  \hat{\delta}({\bi r} ; {\bar{\bi r}}_{k{H}},{{\bi r}}_{k{M}})
{\bi \mu}_k  , 
\ena 
where ${\bar{\bi r}}_{k{H}}$, 
${\bi \xi}_k$,  and ${\bi \mu}_k$  are defined by eqs A2,  A5, and  1, 
respectively. 
 We then find the total polarization 
${\bi M}= \int d{\bi r}{\bi p}({\bi r})=
\sum_k {\bi \mu}_k$. Without ions, the electric field ${\bi E}=-\nabla\Phi$ 
away from the charge positions satisfies 
\be 
\nabla\cdot({\bi E}+ 4\pi{\bi p})=0.
\en  
We  obtain eq 16  by 
multiplying  $z$ to the above relation and integrating 
in the cell.

To calculate  the  average polarization  
$P(z)$ in eq 30,  we 
introduce  the laterally  integrated polarization, 
\be 
M_\perp(z) =\int dxdy~ p_z({\bi r}) . 
\en  
Then, $P(z)=\av{p_z({\bi r})}= 
\av{M_\perp(z)}/L^2$.  The  total polarization along   the $z$ 
axis  in eq 9  is given by $M_z=\int_0^H  dz M_\perp(z)$.  
This  $M_\perp(z)$
 is obtained from eq B3 if  we  replace 
$\hat{\delta} ({\bi r} ; {\bi r}_{1},{\bi r}_2)$  by   the 1D 
symmetrized $\delta$-function, 
\bea
%\hat{\delta}(z; z_1,z_2)
&&  
\int_0^1 d\lambda~ \delta ( z-\lambda z_1  
-(1-\lambda)z_2)\nonumber\\
&&=  [{\theta(z-z_1)-\theta(z-z_2)}]/({z_2-z_1}), 
\ena 
where   $\theta(u)$ is the step function.  We then find   
\bea 
&&\hspace{-1cm}
M_\perp(z)= -{q_H}
\sum_k [ \theta(z- z_{k{H1}})+\theta(z- z_{k{H2}})\nonumber\\
&&\hspace{2cm} -2\theta(z- z_{k{M}})], 
\ena 
which leads to eq 30.

\vspace{2mm}
\noindent{\bf Appendix C: 
Behavior of $e_s(z)$ away from Walls }\\
\setcounter{equation}{0}
\renewcommand{\theequation}{C\arabic{equation}}

Using  eq 29  we discuss the behavior of the 
fluctuation amplitude $e_s(z)$ of the electric field 
due to the surface charge deviations away from the walls. 
For $z\gg \xi_{s0}$ and   
$H-z\gg \xi_{sH}$, we can replace 
  $S_{\lambda  k}$ by $S_{\lambda  0}= 
\lim_{k\to 0}S_{\lambda k}$. 

First, if $L/4\pi<H$, 
we consider the   region  with $z \gs  L/4\pi$ and $H- z \gs  L/4\pi$, 
where we pick up the contributions from 
the smallest  $k=2\pi/L$ to obtain 
\be 
 e_s(z)^2\cong  \frac{32\pi^2  }{L^2}
\sum_{\lambda=0,H} 
S_{\lambda  0} \exp[{-4\pi |z-\lambda|/L}] . 
\en 
In fact, in Fig.3(c), 
the right hand side is $70\%$ of the numerical value at $z=H/2$. 
 
Second, we consider the thin film 
limit  $L/4\pi \gg  H \gg \xi_{s\lambda}$, though this is not the case 
for our cell  with $L=H$.  In this case,    we may 
 replace $2\pi L^{-2} \sum_{{\bi k}\neq{\bi 0}}$ 
by the integral $ \int_{0}^\infty dk k$ (with $k$ being 
continuous) to obtain    
\be 
e_s(z)^2\cong \pi  S_{0 0} /{z^2} 
+  \pi S_{H 0}/{(H-z)^2} .
\en  
From this relation, we estimate $e_s(H/2) 
\sim  0.1 e/H\sigma\sim 1.5 (\sigma/H)$ $ [{\rm V}/{\rm nm}]$  
on the film midplane $z=H/2$, where 
$\sigma$ is  the molecular size$(\sim 3~{\rm \AA})$.

\end{document}